\documentclass[aps,groupedaddress,nofootinbib,floatfix,preprintnumbers]{revtex4}
\usepackage{amsfonts,amscd,amsmath,amssymb}
\usepackage{graphicx,float,xcolor}
\usepackage[section]{placeins}

\def\ep{\text{e}}
\def\g{\mathsf{g}}
\def\oh{\frac{1}{2}}
\def\s{\mathsf{s}}
\def\k{\mathsf{k}}
\def\n{\mathsf{n}}
\def\m{\mathsf{m}}

\def\rv{r_v}

\def\r0{r_0}

\def\rq{r_q}

\def\ro{r_{\text{\tiny 0}}}
\def\vo{{\text v}_{\text{\tiny 1}}}
\def\vp{{\text v}_{\text{\tiny p}}}
\def\vm{{\text v}_{\text{m}}}

\def\lb{\lambda_{\text{b}}}
\def\vd{\text{v}_{\text{d}}}

\def\ld{\lambda_{\text{d}}}
\def\vcd{\text{v}_{\hspace{-1.3pt}{\scriptscriptstyle\times}}}
\def\w{\text w}
\def\vc{\text{v}_c}
\def\vz{{\text v}_{\text{\tiny 0}}}
\def\Qqb{\text{\tiny Q}\bar{\text{\tiny q}}}
\def\QQb{\text{\tiny Q}\bar{\text{\tiny Q}}}
\def\qQb{\text{\tiny q}\bar{\text{\tiny Q}}}
\def\5Q{\text{\tiny 5Q}}
\def\4Q{\text{\tiny 4Q}}
\def\QQq{\text{\tiny QQq}}
\def\3Q{\text{\tiny 3Q}}

\def\QQ{\text{\tiny QQ}}
\def\i{\text{\tiny (I)}}
\def\ii{\text{\tiny (II)}}
\begin{document}
\title{$QQ\bar Q\bar Q$ Quark System and Gauge/String Duality II: String Breaking}
\author{Oleg Andreev}
 \affiliation{L.D. Landau Institute for Theoretical Physics, 142432 Chernogolovka, Russia}
 \begin{abstract} 
This is the second of two companion papers in which we continue developing the construction of the fully heavy tetraquark system using the gauge/string duality. Here, we extend this construction to model $SU(3)$ gauge theory with two light flavors of equal mass.  As an application, we analyze the two lowest Born-Oppenheimer potentials, both analytically and numerically, for rectangular and collinear geometries. Our analysis reveals several critical scales associated with string reconnection and breaking, as well as junction annihilation and pinching. In addition, we find that the string tension is universal, whereas the string breaking distance is not. 
\end{abstract}
\maketitle
\vspace{0.5cm}
\section{Introduction}
\renewcommand{\theequation}{1.\arabic{equation}}
\setcounter{equation}{0}

The quark model \cite{GM,Zw} is entering an era of exotic hadrons, making the longstanding question of how quarks are bound inside hadrons increasingly urgent. An important example is the discovery of several exotic hadrons composed of two charm quarks and two charm antiquarks, known as all-charm tetraquarks \cite{Tcccc}. 

\begin{figure}[htbp]
\centering
\includegraphics[width=11.5cm]{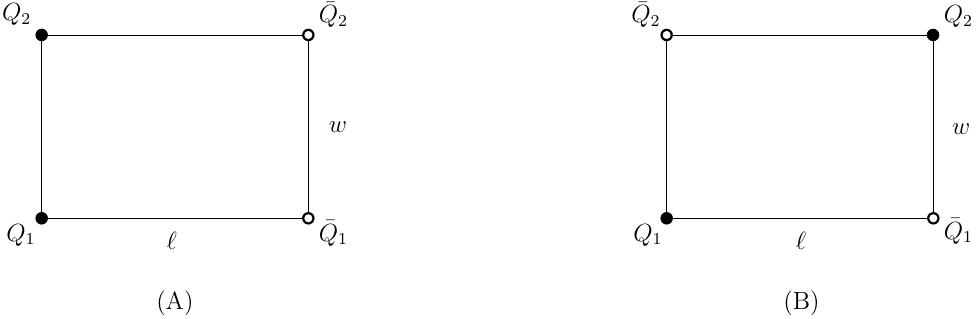}
\caption{{\small A rectangle of length $\ell$ and width $w$. The heavy quarks (antiquarks) are placed at the vertices according to type-A ordering (left) and type-B ordering (right).}}
\label{rect}
\end{figure}

This is the second of two companion papers in which we continue developing the string description of the fully heavy tetraquark system and calculate its first two Born-Oppenheimer (B-O) potentials using the gauge/string duality.\footnote{In fact, this work is part of a broader program aimed at studying fully heavy hadrons. See \cite{az1,a3Q} for discussions of the $Q\bar Q$ and $QQQ$ systems.} In the first paper \cite{a4Q}, we considered the case in which the heavy quarks and antiquarks are placed at the vertices of a rectangle in two different ordering, as shown in Fig.\ref{rect}. We kept the ratio of the length to the width fixed that significantly simplifies the analysis
by reducing the number of parameters by one. Here, we extend the previous analysis in two ways. First, we introduce light quarks, mimicking $SU(3)$ gauge theory with two flavors of equal mass, which allow us to address the issue of string breaking. Second, we keep the width of the rectangle fixed, enabling us to make a contact with the results obtained in lattice QCD \cite{alexa,suga,bicu4Q}.

The outline of this paper is as follows. To make the presentation self-contained, Sec.II briefly reviews the theoretical framework and the specific model in which we will work. Central to this approach is a mixing analysis based on a correlation matrix whose elements give rise to a model Hamiltonian $\mathcal{H}$. The potentials are the eigenvalues of this Hamiltonian. Sec.III is devoted to constructing the string configurations in five dimensions needed to determine the diagonal elements of ${\cal H}$. Here we also introduce the length scales characterizing the transitions between distinct configurations. These length scales are associated with various types of string interactions. Then, in Sec.IV, we discuss the two lowest B-O potentials and compare the results with those of lattice gauge theory. In Sec.V, we comment on a special type of string excitations, the case of collinear geometry, and potential models. We conclude with a discussion in Sec.VI. As in \cite{a4Q}, we include Appendix A, which contains our notation and definitions. However, we do not reproduce the technical results for Nambu-Goto strings in five-dimensional space and instead refer the reader to Appendix B of \cite{a4Q}. In Appendix B, we provide a minimal set of technicalities concerning the $Q\bar Q$ and $QQq$ systems. Finally, in Appendix C, we discuss the five-dimensional analogs of the tetraquark string configuration. 

\section{Preliminaries}
\renewcommand{\theequation}{2.\arabic{equation}}
\setcounter{equation}{0}

In this section we discuss basic features of our construction used to obtain the B-O potentials of the $QQ\bar Q\bar Q$ system.

\subsection{General procedure}

In studying the tetraquark system, we use an approach based on a correlation matrix whose elements give rise to the model Hamiltonian
\begin{equation}\label{H}
{\cal H}=
\begin{pmatrix}
\,E_1& {} & \Theta_{ij}  \\
\,{} & \ddots & {}\,\\
\,\,\Theta_{ij}  & {} & E_n\,
\end{pmatrix}
\,.
\end{equation}
Here the diagonal elements represent the energies of the string configurations and the off-diagonal elements describe the strength of mixing between these configurations. In general, $n$ is infinite, but in practice it is taken to be finite, depending on the purpose of the calculation. The two lowest potentials, $V_0$ and $V_1$, are given by the two smallest eigenvalues of ${\cal H}$. 

Now consider string configurations for the $QQ\bar Q\bar Q$ system from the standard viewpoint in four dimensions \cite{XA}. We restrict ourselves to $N_f=2$, i.e., two dynamical flavors of equal mass, although the extension to full QCD with $N_f=2+1$ is straightforward. 

\begin{figure}[htbp]
\centering
\includegraphics[width=15cm]{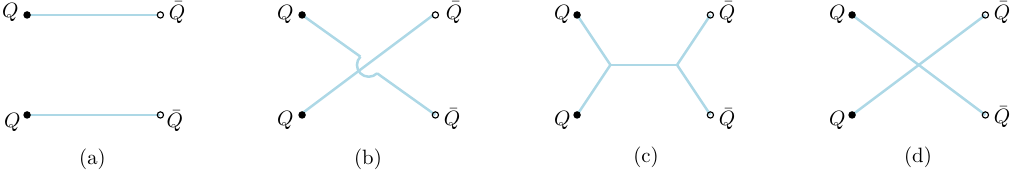}
\caption{{\small Four-dimensional string configurations without light quarks at the string endpoints: (a)-(b) two-meson configurations, (c) a tetraquark configuration, and (d) a pinched tetraquark configuration. Since configurations (c) and (d) describe the genuine four-body interactions of quarks, we call them the tetraquark configurations.}}
\label{c4A}
\end{figure}

\begin{figure}[htbp]
\centering
\includegraphics[width=11.75cm]{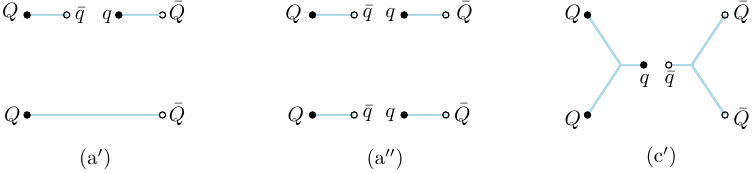}
\caption{{\small Some configurations with light quarks: ($\mathrm{a}'$)-($\mathrm{a}''$) multimeson configurations and ($\mathrm{c}'$) a two-baryon configuration. Here and below, we label them by the same letters as the parent configurations of Fig.\ref{c4A}, from which they  are obtained by adding virtual quark-antiquark pairs. The number of primes is equal to the number of such pairs.}}
\label{c4Aq}
\end{figure}

For type-A ordering, it is easy to guess the simplest disconnected configurations \cite{a4Q}. They are the two-meson configurations (a) and (b) shown in Fig.\ref{c4A}. 
Each consists of the heavy quark and antiquark sources connected by strings and looks like a pair of heavy mesons. To proceed further, it is helpful to think of, to some extent, the string junctions and light quark pairs as elementary excitations. Adding a pair of junctions gives the tetraquark and pinched tetraquark configurations \cite{a4Q}, namely configurations (c) and (d) in Fig.\ref{c4A}. On the other hand, adding light quarks yields the disconnected string configurations shown in Fig.\ref{c4Aq}.
 
\begin{figure}[H]
\centering
\includegraphics[width=11cm]{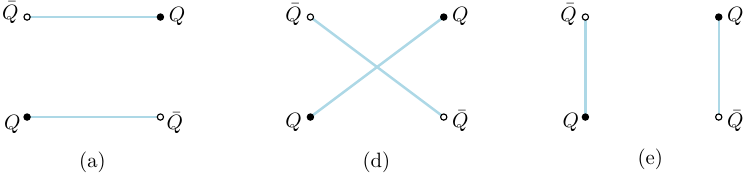}
\caption{{\small Four-dimensional string configurations without light quarks: (a) and (e) two-meson configurations, and (d) a pinched tetraquark configuration.}}
\label{c4B}
\end{figure}

\begin{figure}[H]
\centering
\includegraphics[width=7.5cm]{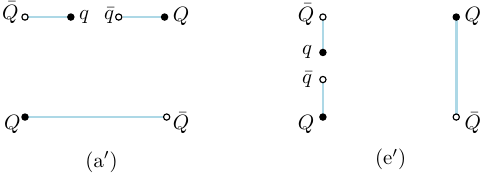}
\caption{{\small Multimeson configurations arising from the two-meson configurations of Fig.\ref{c4B}.}}
\label{c4Bq}
\end{figure}

Similarly, for type-B ordering, there are the two-meson configurations and the pinched tetraquark configuration, as shown in Fig.\ref{c4B} \cite{a4Q}. In addition, string breaking gives rise to multimeson configurations. These configurations are 
\noindent shown in Fig.\ref{c4Bq}. It is noteworthy that, in spite of the different ordering, configurations $(\mathrm{a})$, $(\mathrm{a}')$, and (d) in the above 
\noindent figures are in fact identical. 

We conclude this discussion with a few remarks. First, each disconnected configuration can be interpreted as a hadronic molecule composed of mesons or baryons. However, we suppose that its binding energy is negligible so that the energy equals the sum of the rest energies of the constituent hadrons, as in lattice QCD \cite{alexa,suga,bicu4Q}. Second, we keep both configuration (c) and (d) because their energies differ at large $\ell$  \cite{a4Q} (also see Fig.\ref{E01054s}). Third, we try to consider only configurations relevant for the two low-lying B-O potentials. Finally, there is no tetraquark configuration for type-B ordering, also in five dimensions \cite{a4Q}. 

\begin{figure}[htbp]
\centering
\includegraphics[width=13.5cm]{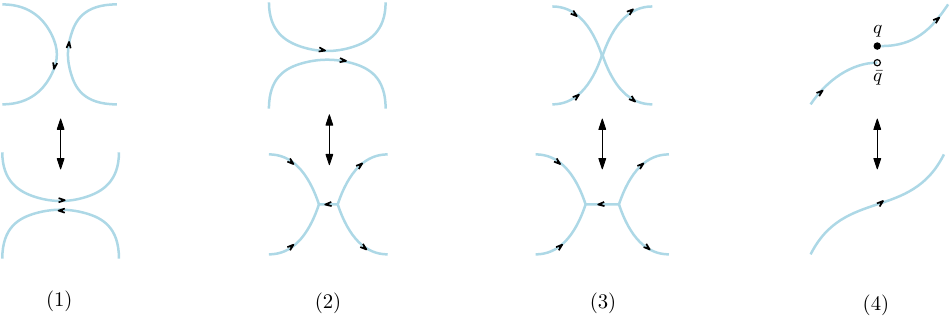}
\caption{{\small Some string interactions: (1) reconnection, (2) junction annihilation, (3) pinching, and (4) string breaking.}}
\label{stri}
\end{figure}

Transitions between configurations can be interpreted in terms of string interactions. In Fig.\ref{stri}, we sketch four types 
\noindent of interactions relevant to the discussion in the following sections, where we will introduce the notion of a critical length to characterize the transitions quantitatively. Note that pinching has a transparent interpretation in five dimensions as the creation of a brane-antibrane bound state.

\subsection{The five-dimensional model}

For convenience, we include a very brief summary.\footnote{We refer the reader to \cite{a4Q} for further details.} It is as follows. The five-dimensional geometry is chosen to be a one-parameter deformation of Euclidean $\text{AdS}_5$ space of radius $R$:

\begin{equation}\label{metric}
ds^2=\ep^{\s r^2}\frac{R^2}{r^2}\Bigl(dt^2+(dx^i)^2+dr^2\Bigr)
\,.
\end{equation}
Here $r$ denotes the fifth (radial) coordinate, while $\s$ is a deformation parameter. The boundary is located at $r=0$, and the soft wall at $r=1/\sqrt{\s}$. This wall confines strings near the boundary.

A fundamental string living in this five-dimensional space is governed by the Nambu-Goto action 

\begin{equation}\label{ng}
S_{\text{\tiny NG}}=\frac{1}{2\pi\alpha'}\int d^2\xi\,\sqrt{\gamma^{(2)}}
\,,
\end{equation}
where $\gamma$ is an induced metric, $\alpha'$ is a string parameter, and $\xi^i$ are world-sheet coordinates. 

A baryon vertex, which is the high-dimensional analogue of a string junction,\footnote{We use this terminology to distinguish between the four-dimensional string junction and its five-dimensional counterpart.} is a point-like object whose action in the static gauge takes the form

\begin{equation}\label{baryon-v}
S_{\text{\tiny V}}=\tau_v\int dt \,\frac{\ep^{-2\s r^2}}{r}
\,,
\end{equation}
with $\tau_v$ a free parameter. At zero baryon chemical potential, it is natural to assume that the antibaryon vertex is described by the same action, so that $S_{\bar{\text{\tiny V}}}=S_{\text{\tiny V}}$.

The light quarks at string endpoints in the bulk are introduced by adding to the world-sheet action a boundary term, namely the action of a massive particle propagating in a curved space. For the geometry \eqref{metric}, this action, written in static gauge, is 

\begin{equation}\label{Sq}
S_{\text q}=\m\int dt \frac{\ep^{\frac{1}{2}\s r^2}}{r}
\,,
\end{equation}
where $\m$ is a dimensionless parameter.

To make the following discussion concrete, we need to specify the model parameters. We use the values suggested in \cite{a-strb, astbr3Q} for the case of the $Q\bar Q$ and $QQQ$ quark systems in the presence of light quarks, namely $\s=0.45\,\text{GeV}^2$, $\g=\frac{R^2}{2\pi\alpha'}=0.176$, $\k=\frac{\tau_v}{3\g}=-0.102$, and $\n=\frac{\m}{\g}=3.057$.\footnote{Unlike \cite{a4Q}, where $\k=-0.083$ is used for pure $SU(3)$ gauge theory, we choose $\k=-0.102$. This choice is motivated by the absence of lattice data for the three-quark potential in QCD with light quarks and by the phenomenological rule $E_{\QQ}(\ell)=\oh E_{\QQb}(\ell)$ as $\ell\to 0$.} These values will be used in all subsequent estimates. Importantly, no additional parameters are introduced to describe the tetraquark system.

\section{String configurations in five dimensions}
\renewcommand{\theequation}{3.\arabic{equation}}
\setcounter{equation}{0}

Now we turn  to the string configurations in the geometry \eqref{metric}. As said previously, we keep the rectangular width fixed by setting 

\begin{equation}\label{gc}
	w=const
	\,,
\end{equation}
which we call the geometrical constraint. 

One way to develop some intuition for these objects is to place their four-dimensional counterparts on the boundary of five-dimensional space. A gravitational force then pulls the light quarks (antiquarks) and strings into the interior, whereas the heavy quarks and antiquarks remain at rest. The analysis proceeds straightforwardly along the lines of \cite{a4Q}. Only a few technical details need to be explained, mainly because of the constraint \eqref{gc}.\footnote{In \cite{a4Q}, we used the constraint $w/\ell=const$.}

\subsection{Type-A ordering}
\subsubsection{The two-meson configurations}

Each two-meson configuration consists of two parts, each of which is shown in Fig.\ref{5QQb}a. Given our assumption, its energy is the sum of the energies of these parts. 

For configuration (a), the total energy as a function of $\ell$ is simply

\begin{equation}\label{a}
E^{\text{(a)}}=2E_{\QQb}(\ell)
\,,	
\end{equation}
where $E_{\QQb}$ and $\ell$ are given parametrically by Eqs.\eqref{EQQb}.

It follows immediately from the formulas of Appendix B that the small-$\ell$ expansion takes the form 

\begin{equation}\label{Ela-small}
E^{\text{(a)}}(\ell)=-\frac{\alpha^{\text{(a)}}}{\ell}+4c+o(1)
\,,\qquad\text{with}\qquad
\alpha^{\text{(a)}}=2\alpha_{\QQb}
\,.
\end{equation}
Here the coefficient $\alpha_{\QQb}$ is given by \eqref{alpha-QQb}, and $c$ is the normalization constant. For large $\ell$, the expansion is 

\begin{equation}\label{Ela-large}
E^{\text{(a)}}(\ell)=
2\sigma\ell+C^{\text{(a)}}
+o(1)
\,,\qquad\text{with}\qquad
C^{\text{(a)}}=2C_{\QQb}
\,.
\end{equation}
Here $\sigma$ and $C_{\QQb}$ are defined in \eqref{EQQb-large}. Importantly, $\sigma=\ep\g\s$ is the physical string tension whose universality was demonstrated for the $Q\bar Q$, $QQQ$, and $QQ\bar Q\bar Q$ systems \cite{az1,a3Q, a4Q}.

For configuration (b), the total energy is  

\begin{equation}\label{Eb}
E^{\text{(b)}}=2E_{\QQb}(\sqrt{\ell^2+w^2})
\,.
\end{equation}
In parametric form, it is given by 

\begin{equation}\label{Eb2}
	\ell=\sqrt{\frac{4}{\s}\Bigl({\cal L}^+(\lambda)\Bigr)^2-w^2}\,,
	\qquad
	E^{\text{(b)}}=4\g\sqrt{\s}\,{\cal E}^+(\lambda)+4c
	\,,
\end{equation}
as follows from \eqref{EQQb}. Here the functions ${\cal L}^+$ and ${\cal E}^+$ are defined in Appendix A. The parameter $\lambda$ varies from $\lb$ to $1$, where $\lb$ is a solution of the equation $w=\frac{2}{\sqrt{s}}{\cal L}^+(\lambda)$ in the interval $[0,1]$. 

The corresponding asymptotic expansions can be obtained from the expressions \eqref{Eb2}. For small $\ell$, the expansion reads

\begin{equation}\label{Eb-small}
E^{\text{(b)}}=c^{\text{(b)}}+B\ell^2+o(\ell^2)
\,,
\end{equation}
where $c^{\text{(b)}}=4\g\sqrt{\s}\,{\cal E}^+(\lb)+4c$ and 
$B=\frac{\g\s}{w}\frac{\partial_\lambda{\cal E}^+}{\partial_\lambda{\cal L}^+}\bigl\vert_{\lambda=\lb}$. Similarly, for large $\ell$, it reads 

\begin{equation}\label{Eb-large}
E^{\text{(b)}}(\ell)=2\sigma\ell+C^{\text{(b)}}
+o(1)
\,,
\qquad\text{with}\qquad
C^{\text{(b)}}=2C_{\QQb}
\,.
\end{equation}
At this order, the large-$\ell$ expansions of configurations (a) and (b) coincide. However, the corrections matter: for nonzero $w$, configuration (b) always has a higher energy than configuration (a). The reason is as follows. Both functions ${\cal L}^+(\lambda)$ and ${\cal E}^+(\lambda)$ increase monotonically in the interval $[0,1]$. For a given $\ell$, the value of $\lambda^{\text{(b)}}$ must be larger than that of $\lambda^{\text{(a)}}$ because of the shift by $w$ in \eqref{Eb2}. Hence, ${\cal E}^+(\lambda^{\text{(b)}})>{\cal E}^+(\lambda^{\text{(a)}})$. This implies that configuration (b) is irrelevant for the ground state potential and may contribute only to higher B-O potentials. 
 
\subsubsection{The tetraquark configuration}

Like in \cite{a4Q}, the tetraquark configuration is constructed from the basic configurations (I) and (II) described in Appendix C. Its energy is written parametrically as:

\begin{equation}\label{Elc1}
\ell=\begin{cases}
\ell^{\,\ii} (v)\,,\\ 
\ell^{\,\i}(v)\,\,,
\end{cases}
\quad
E^{\text{(c)}}=\begin{cases}
E^{\,\ii} (v)\,,\,& \,\vcd \leq v\leq \vz\,\,,\\
E^{\,\i}(v)\,\,,\,\,\,&\,\vz\leq v\leq \vo\,
\end{cases}
\end{equation}
if $w\leq\w$ and 
\begin{equation}\label{Elc2}
\ell=\ell^{\,\ii}(v) \,,\qquad
E^{\text{(c)}}=E^{\,\ii}(v) 
\,,
\qquad
\,\vcd \leq v\leq \vo
\,
\end{equation}
if $w\geq\w$. Here $E^{\,\i}$ and $E^{\,\ii}$ denote the energies of the basic configurations defined in Eqs.\eqref{EIl} and \eqref{ElII}. The bounds on $v$ depend on the width $w$ and are defined in Appendix C. Their meaning is as follows: $\vcd$ is the transition point between configurations (c) and (d), $\vz$ is the transition point between the basic configurations (I) and (II), and $\vo$ corresponds to the large-$\ell$ limit.\footnote{For consistency, they must satisfy $\vcd\leq\vz\leq\vo$. A typical example is $\vcd=0.223$, $\vz=0.267$, and $\vo=0.343$ at $w=0.19\,\text{fm}$.} The value of $\w$ is determined by the equation $\vz(w)=\vo(w)$. Numerically, $\w=0.779\,\text{fm}$.

Since $\ell$ is bounded from below, it is meaningless to discuss the small-$\ell$ behavior of $E^{\text{(c)}}$ unless $\ell^{\ii}(\vcd)$ is sufficiently small. The large-$\ell$ expansion, however, is meaningful and reads

\begin{equation}\label{Ec-large}
E^{\text{(c)}}=
\sigma\ell
+
C^{\text{(c)}}+o(1)
\,,
\qquad
C^{\text{(c)}}=
\begin{cases}
	\g\sqrt{\s}\Bigl(4{\cal E}^+(\alpha,\vo)-2\ep\sin\beta{\cal L}^+(\alpha,\vo)-{\cal J}(\vo)
+6\k\frac{\ep^{-2\vo}}{\sqrt{\vo}}\Bigr)
+4c\,& \text{if}\quad w\leq\w\,,\\
\g\sqrt{\s}\Bigl(4{\cal E}^-(\lambda,\vo)-2\ep\sin\beta{\cal L}^-(\lambda,\vo)-{\cal J}(\vo)
+6\k\frac{\ep^{-2\vo}}{\sqrt{\vo}}\Bigr)
+4c
\, & \text{if}\quad w\geq\w
\,,
\end{cases}
\end{equation}
as follows from Eqs.\eqref{EIdi5} and \eqref{EIIdi5}. The angles $\alpha$ and $\beta$, as well as the parameter $\lambda$, are defined below these equations. The functions $\mathcal{E}$, $\mathcal{L}$, and $\mathcal{J}$ are given in Appendix A. Here, $\sigma$ is the same string tension as in the $Q\bar Q$, $QQQ$, and $QQq$ systems \cite{az1,a3Q,aQQq}. Thus, it is universal. 

The important fact is that $\vcd$, $\vz$, and $\vo$ vanish in the limit $w\to 0$, while preserving the ordering $\vcd<\vz<\vo$. This allows one to consider the diquark limit by taking $w\to 0$ and $\ell\to\infty$. In this limit, using the expansions \eqref{diq-limit}, \eqref{Ec-large} can be rewritten as 

\begin{equation}\label{diquarkEc}
E^{\text{(c)}}=	E_{\QQb}(\ell)+2 E_{\QQ}(w)\,,
\end{equation}
where 
\begin{equation}\label{diquarkEc2}
	E_{\QQb}(\ell)=\sigma\ell+C_{\QQb}+o(1)\,,\qquad
	E_{\QQ}(w)=-\frac{\alpha_{\QQ}}{w}+c+\sigma_0 w+o(w)
	\,.
\end{equation}
Here $\alpha_{\QQ}=-w_0 E_0\g$ and $\sigma_0=\frac{1}{w_0}\bigl(E_1+\frac{w_1}{w_0}E_0\bigr)\g\s$, with the coefficients defined in Appendix C. These expressions are consistent with heavy quark-diquark symmetry \cite{wise}.  We also note that the expression for the heavy quark-quark potential $E_{\QQ}$ is the same as that obtained from the $QQQ$ system in \cite{a3Q} and from the $QQq$ system in \cite{aQQq}. 

\subsubsection{The pinched tetraquark configuration}

\begin{figure}[htpb]
\centering
\includegraphics[width=7.2cm]{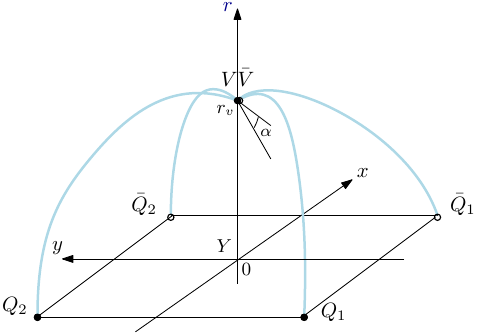}
\caption{{\small A pinched tetraquark configuration in five dimensions. The baryon vertices are located at the same point on the $r$-axis, at $r=\rv$, and their projection onto the boundary lies at the origin. By symmetry, all tangent angles at $r=\rv$ are equal. We denote them by $\alpha$.}}
\label{pinched5d}
\end{figure}

We now consider the pinched tetraquark configuration in five dimensions. One might expect it to arise as a special limit of the tetraquark configuration, but this is not the case. Both configurations can exist independently over a wide range of lengths, as seen from  Fig.\ref{E01054s}. Configuration (d) is shown in Fig.\ref{pinched5d}. In \cite{a4Q}, it was studied by keeping the ratio 
between the length and the width fixed. Here we extend the analysis to the case in which the width is kept fixed. 

The total action governing this configuration is the sum of the Nambu-Goto actions of the four strings and the actions of the two vertices, namely $
S=\sum_{i=1}^4 S_{\text{\tiny NG}}^{(i)}\,+2S_{\text{\tiny V}}$. Extremizing it with respect to $\rv$ yields the $r$-component of the force balance equation at this point \cite{a4Q}, from which the tangent angle $\alpha$ can be expressed in terms of $v$ as 

\begin{equation}\label{fbe-d}
	\sin\alpha=\frac{3}{2}\k(1+4v)\ep^{-3v}
	\,.
\end{equation}
Note that $\alpha$ is negative because $\k<0$.

We proceed further as in \cite{a4Q}. Using the formulas of Appendix B of \cite{a4Q} for the string lengths and energies, and combining them with \eqref{baryon-v}, we obtain

\begin{equation}\label{Eld}
	\ell=\sqrt{\frac{4}{\s}\Bigl({\cal L}^-(\lambda,v)\bigr)^2-w^2}\,,
	\qquad
	E^{\text{(d)}}=2\g\sqrt{\s}\biggl(2{\cal E}^-(\lambda,v)+3\k\frac{\ep^{-2v}}{\sqrt{v}}\biggr)+4c
	\,.
\end{equation}
Here $\lambda$ is given by $\lambda=-\text{ProductLog}(-v\,\ep^{-v}/\cos\alpha)$, where $\text{ProductLog}(z)$ denotes the principal solution for $u$ in the equation $z =u\ep^u$ \cite{wolf}. Together with \eqref{fbe-d}, Eqs.\eqref{Eld} define a parametric form of $E^{\text{(d)}}(\ell)$. The parameter $v$ varies from $\vd$ to $\vp$, both of which lie in the interval $[0,1]$. The lower bound, which depends on $w$, is determined from the equation $w=\frac{2}{\sqrt{s}}{\cal L}^-(\lambda,v)$ and corresponds to the small-$\ell$ limit. On the other hand, the upper bound, which is independent of $w$, is determined from the equation \cite{a4Q}

\begin{equation}\label{v1p}
	1-v^2\ep^{2(1-v)}-\frac{9}{4}\k^2(1+4v)^2\ep^{-6v}=0
	\,,
\end{equation}
and corresponds to the large-$\ell$ limit. Numerically, $\vp=0.959$, whereas $\vd=0.177$ at $w=0.19\,\text{fm}$ and $\vd=0.481$ at $w=0.54\,\text{fm}$, for example. A short inspection shows that $\ell(v)$ is an increasing function of $v$ that vanishes at $v=\vd$ and diverges at $v=\vp$. 

To analyze the small-$\ell$ behavior of $E^{\text{(d)}}$, we need the expansions of $\ell(v)$ and $E^{\text{(d)}}(v)$ near $v=\vd$. Keeping only the two leading terms, we get 

\begin{equation}\label{Ed-small}
E^{\text{(d)}}=c^{\text{(d)}}+D\ell^2
+o(\ell^2)
\,,
\end{equation}
where
$c^{\text{(d)}}=2\g\sqrt{\s}\Bigl(2{\cal E}^-(\ld,\vd)+3\k\frac{\ep^{-2\vd}}{\sqrt{\vd}}\Bigr)+4c$ and $D=\frac{\g\s}{w}(\partial_v{\cal E}^--\frac{3}{4}\k(1+4v)v^{-\frac{3}{2}}\ep^{-2v})/\partial_v{\cal L}^-\bigl\vert_{v=\vd}$. Here, $\ld$ is shorthand for $\lambda(\vd)$.

In a similar spirit, we can analyze the large-$\ell$ behavior. In the limit $v\to\vp$, or equivalently $\lambda\to1$, the strings become infinitely large, as follows from Eq.\eqref{L-y=1}. At leading order, Eqs.\eqref{Eld} reduce to 

\begin{equation}\label{Eld-large}
	\ell=-\frac{2}{\sqrt{\s}}\ln(1-\lambda)+O(1)\,,
\qquad
	E^{\text{(d)}}=-4\ep\g\sqrt{\s}\ln(1-\lambda)+O(1)
\,,
\end{equation}
where in the last step we used \eqref{E-y=1}. From this, it follows immediately that 

\begin{equation}\label{Eld-large2}
E^{\text{(d)}}=2\sigma\ell+O(1)
\,,
\end{equation}
as expected. To compute the next-to-leading term, consider 

\begin{equation}\label{Eld-large3}
E^{\text{(d)}}-2\sigma\ell
= 
2\g\sqrt{\s}\Bigl(
2\bigl({\cal E}^-(\lambda,v)-\ep {\cal L}^-(\lambda,v)\bigr)+3\k\frac{\ep^{-2v}}{\sqrt{v}}
\Bigr)+4c+\dots
\,,
\end{equation}
where the dots are subleading by powers of $1/{\cal L}^-$ as $\lambda\to 1$. The large-$\ell$ expansion of the energy is obtained by taking the limit $v\to\vp$ with the help of  \eqref{ILE}. Explicitly, 

\begin{equation}\label{Eld-large4}
E^{\text{(d)}}=2\sigma\ell +C^{\text{(d)}}+o(1)
\,,\qquad\text{with}\qquad 
C^{\text{(d)}}=4c-2\g\sqrt{\s}\Bigl(2{\cal I}(\vp)-3\k\frac{\ep^{-2\vp}}{\sqrt{\vp}}\Bigr)
\,.
\end{equation}
Although, to this order, it coincides with the asymptotic expansion found in \cite{a4Q} for the geometrical constraint $w/\ell=const$, the higher order corrections differ.
 
\subsubsection{What we have learned so far}

One lesson to be learned from concrete examples is that we have considered more string configurations than necessary.\footnote{Here, we retain configuration (b) since its energy only slightly exceeds that of configuration (d).} Another is that, in the absence of disconnected configurations, the insights gained may also apply to the case of pure gauge theory.

For the concrete examples, we consider widths in the range $0.19\,\text{fm}\leq w\leq 0.54\,\text{fm}$. This range is of physical interest because it is compatible with the lattice calculations \cite{suga,bicu4Q}, where $w$ takes values bounded by this range, as well as with phenomenological estimates of the size of a $[bb]$ diquark, which yield $w=0.19\,\text{fm}$ \cite{quigg} and $w=0.37\,\text{fm}$ \cite{deb}.

\begin{figure}[htpb]
\centering
\includegraphics[width=8.75cm]{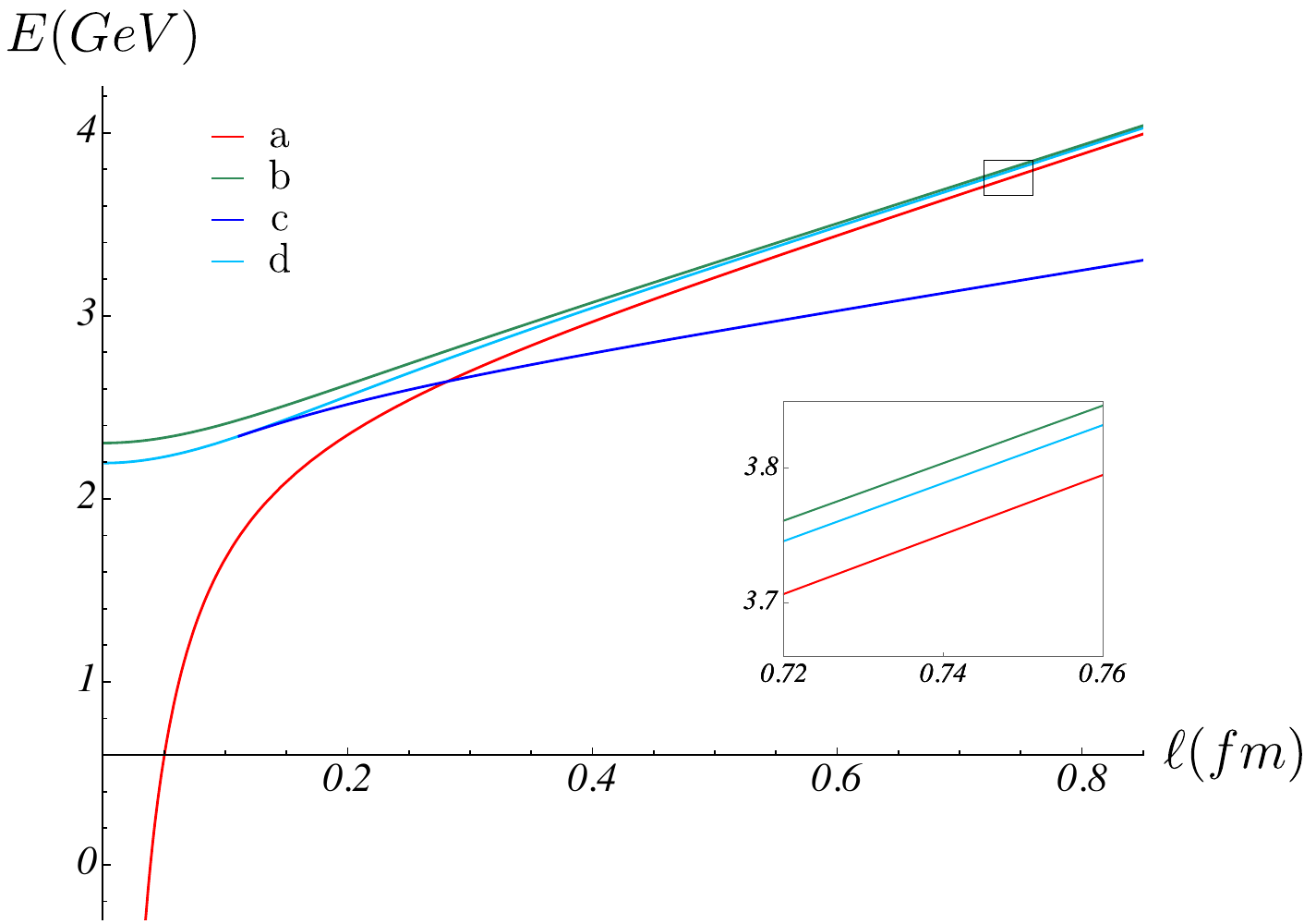}
\hspace{0.3cm}
\includegraphics[width=8.55cm]{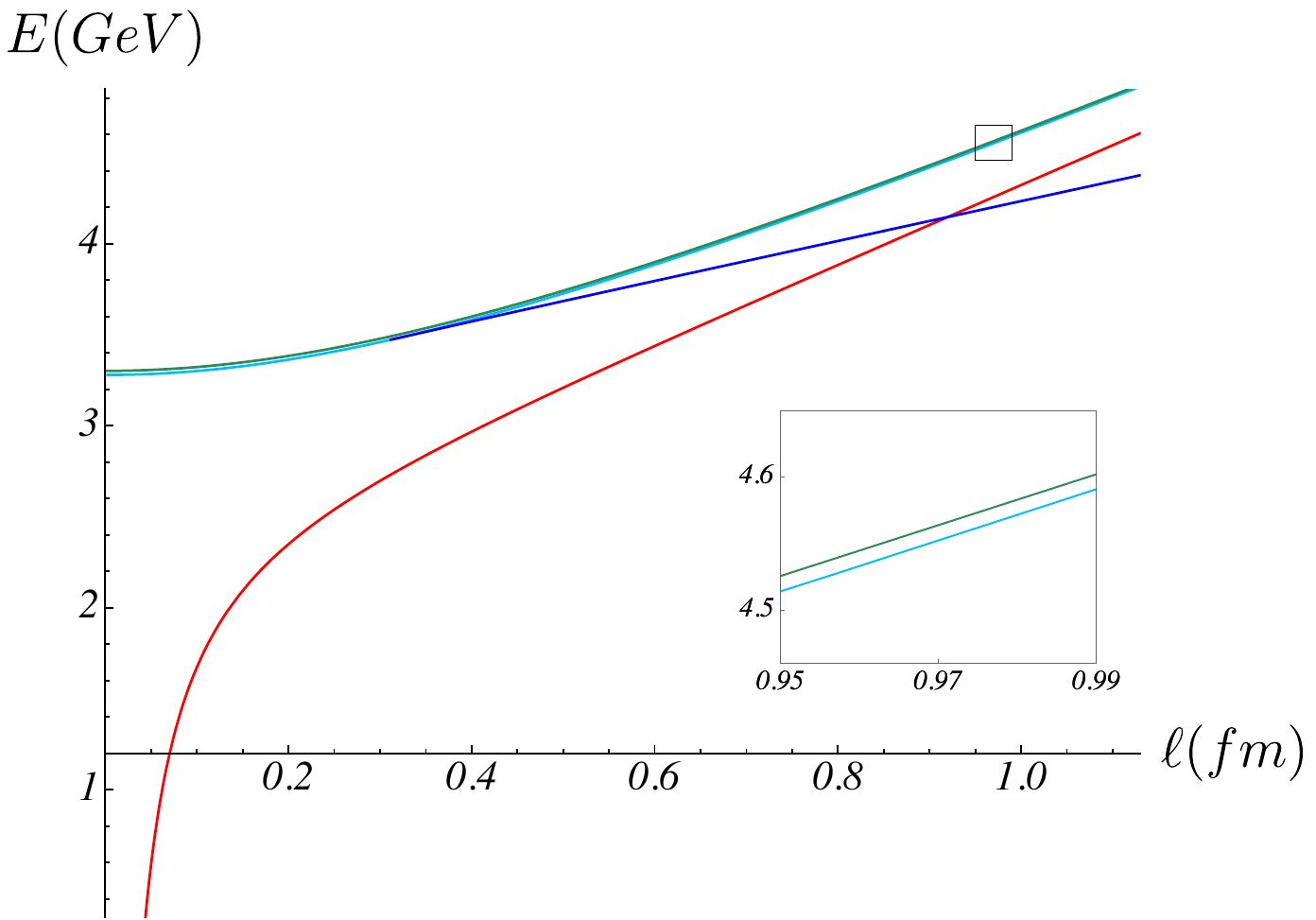}
\caption{{\small The energies of the string configurations (a)-(d) at $w=0.19\,\text{fm}$ (left) and at $w=0.54\,\text{fm}$ (right). Here and below, we set $c=0.623\,\text{GeV}$.}}
\label{E01054s}
\end{figure}

In what follows, we restrict to small and intermediate values of $\ell$. In the left panel of Fig.\ref{E01054s} we plot the 
energies of the string configurations as functions of $\ell$ for $w=0.19\,\text{fm}$. An interesting consequence of the pattern shown in this figure, as well as in others, is the emergence of several characteristic scales that separate different configurations, or, equivalently, different descriptions. To make this more quantitative, we define critical lengths by equating the energies of configurations (i) and (j),

\begin{equation}\label{lijc}
E^{(\mathrm{i})}(\ell)=E^{(\mathrm{j})}(\ell)
\,,
\end{equation}
such that the corresponding critical length is a solution of this equation. Here the indices are used to denote the string configurations. At this value of $\ell$, a transition between these configurations occurs. Such a transition arises from the string interactions of Fig.\ref{stri}. In what follows, we give details illustrating how this works for configurations (c) and (c'), while only mentioning specific results for the others.

The ground state potential is determined by two configurations, so that $V_0=\min\{E^{\text{(a)}},E^{\text{(c)}}\}$. Configuration (a) dominates at small $\ell$ and then transforms to configuration (c) at larger $\ell$. The transition occurs at $\ell=0.282\,\text{fm}$ and can be interpreted as the creation of a pair of baryon vertices. This suggests that the ground state is a mixed state of a hadronic molecule, composed of two mesons, and a tetraquark state.\footnote{In general, such a tetraquark state may be non-compact.} The first excited potential is determined by three configurations, namely $V_1=\min\{E^{\text{(a)}},E^{\text{(c)}},E^{\text{(d)}}\}$.\footnote{First, one has to take the minimum that determines $V_0$.} Configuration (d) dominates at small $\ell$, then configuration (c) becomes dominant at $\ell=0.111\,\text{fm}$ due to unpinching, and, finally, configuration (a) dominates for $\ell>0.282\,\text{fm}$.

One important observation for the subsequent analysis is the following. At large $\ell$, the plots of $E^{\text{(a)}}$, $E^{\text{(b)}}$, and $E^{\text{(d)}}$ seem indistinguishable, since the differences between them become extremely small. As explained in section 1, $E^{\text{(b)}}>E^{\text{(a)}}$, and therefore configuration (b) does not contribute to $V_1$. To understand what happens with $E^{\text{(a)}}$ and $E^{\text{(d)}}$, it is helpful to look at the large-$\ell$ asymptotics \eqref{Ela-large} and \eqref{Eld-large4}. Although their leading terms are the same, their constant terms differ. A simple estimate gives $C^{\text{(a)}}-C^{\text{(d)}}=10\,\text{MeV}$. This implies that for sufficiently large $\ell$ configuration (a) is no longer relevant for $V_1$.  It is straightforward to find the intersection point of the corresponding curves, which lies at approximately $\ell=2.641\,\text{fm}$. This is well above the typical scale of $1\,\text{fm}$ at which string breaking occurs. Therefore, to understand the large-$\ell$ behavior of $V_1$, it suffices to consider disconnected configurations obtained from configuration (a) rather than from (d). Moreover, this remains true as $w$ increases. For example, for $w=0.39\,\text{fm}$ the intersection point is at $\ell=3.348\,\text{fm}$. 

To complete the picture, the right panel of the figure presents the results for $w=0.54\,\text{fm}$. The overall picture remains the same, but with two noteworthy features. First, the difference between $E^{\text{(b)}}$ and $E^{\text{(d)}}$ becomes smaller. Nevertheless the inequality $E^{\text{(b)}}>E^{\text{(d)}}$ still holds and can be checked numerically. A heuristic argument in favor of this is that the contribution of the baryon vertices to the energy of configuration (d) is negative, whereas the contributions of the strings to both energies are nearly the same (see Fig.\ref{c4A}). Second, in addition to the intersection point between $E^{\text{(a)}}$ and $E^{\text{(d)}}$, the other points also shift to larger values of $\ell$. In other words, the critical lengths increase with $w$. For example, the intersection point of $E^{\text{(a)}}$ and $E^{\text{(c)}}$ shifts to $\ell=0.920\,\text{fm}$. 

In short, for lengths below $1\,\text{fm}$ the two lowest potentials, without contributions of light quarks, can be written as $V_0=\min\{E^{\text{(a)}},E^{\text{(c)}}\}$ and $V_1=\min\{E^{\text{(a)}},E^{\text{(c)}},E^{\text{(d)}}\}$. 

\subsubsection{The multimeson and two-baryon configurations}

We begin with the multimeson configurations. As explained in Appendix B, in the case of the $Q\bar Q$ system, string breaking leads to the decay $Q\bar Q\to Q\bar q +q\bar Q$, which in turn gives rise to two possible decay channels for configuration (a): $2Q\bar Q\to Q\bar Q+Q\bar q+q\bar Q$ and $2Q\bar Q\to 2Q\bar q+2q\bar Q$. The corresponding string configurations are shown in Figs.\ref{c4Aq}$\mathrm{a}'$-$\mathrm{a}''$. Given our usual assumption for the energies of the disconnected configurations, their analysis proceeds in an obvious way.

For configuration ($\mathrm{a}'$), we have 

\begin{equation}\label{Eap}
	E^{\text{($\mathrm{a}'$)}}(\ell)
	=E_{\QQb}(\ell)+ 2E_{\Qqb}
	\,.
\end{equation}
Here we assume that at zero baryonic chemical potential $E_{\Qqb}=E_{\qQb}$, with $E_{\Qqb}$ given by \eqref{Qqb}. The small- and large-$\ell$ expansions of $E^{(\mathrm{a}')}$ read 

\begin{equation}\label{Eap-small}
E^{\text{($\mathrm{a}'$)}}(\ell)=-\frac{\alpha_{\QQb}}{\ell}+2c+2E_{\Qqb}+o(1)
\end{equation}
and 
\begin{equation}\label{Eap-large}
E^{\text{($\mathrm{a}'$)}}(\ell)=
\sigma\ell+C^{(\text{a}')}+o(1)
\,,\qquad\text{with}\qquad 
C^{\text{($\mathrm{a}'$)}}=C_{\QQb}+2E_{\Qqb}
\,,
\end{equation}
where we used \eqref{EQQb-small} and \eqref{EQQb-large}.

For configuration ($\mathrm{a}''$), we simply obtain

\begin{equation}\label{Eapp}
	E^{\text{($\mathrm{a}''$)}}=4E_{\Qqb}
	\,,
\end{equation}
so that its energy is independent of $\ell$.

Since the width $w$ is kept fixed, only the internal string of configuration (c)  becomes infinitely long as $\ell\to\infty$. When it breaks, the configuration decays into two doubly heavy baryons, $QQq$ and $\bar Q\bar Q\bar q$, as shown in Fig.\ref{c4Aq}($\mathrm{c}'$). For this configuration, the energy is independent of $\ell$ and is given by   

\begin{equation}\label{Ecp}
	E^{\text{($\mathrm{c}'$)}}=2E_{\QQq}(w)
	\,,
\end{equation}
where $E_{\QQq}$ is the energy of the $QQq$ system. Its dependence on the separation between the heavy quarks is explained in Appendix B. 

\subsubsection{Putting it all together}

\begin{figure}[htpb]
\centering
\includegraphics[width=8.8cm]{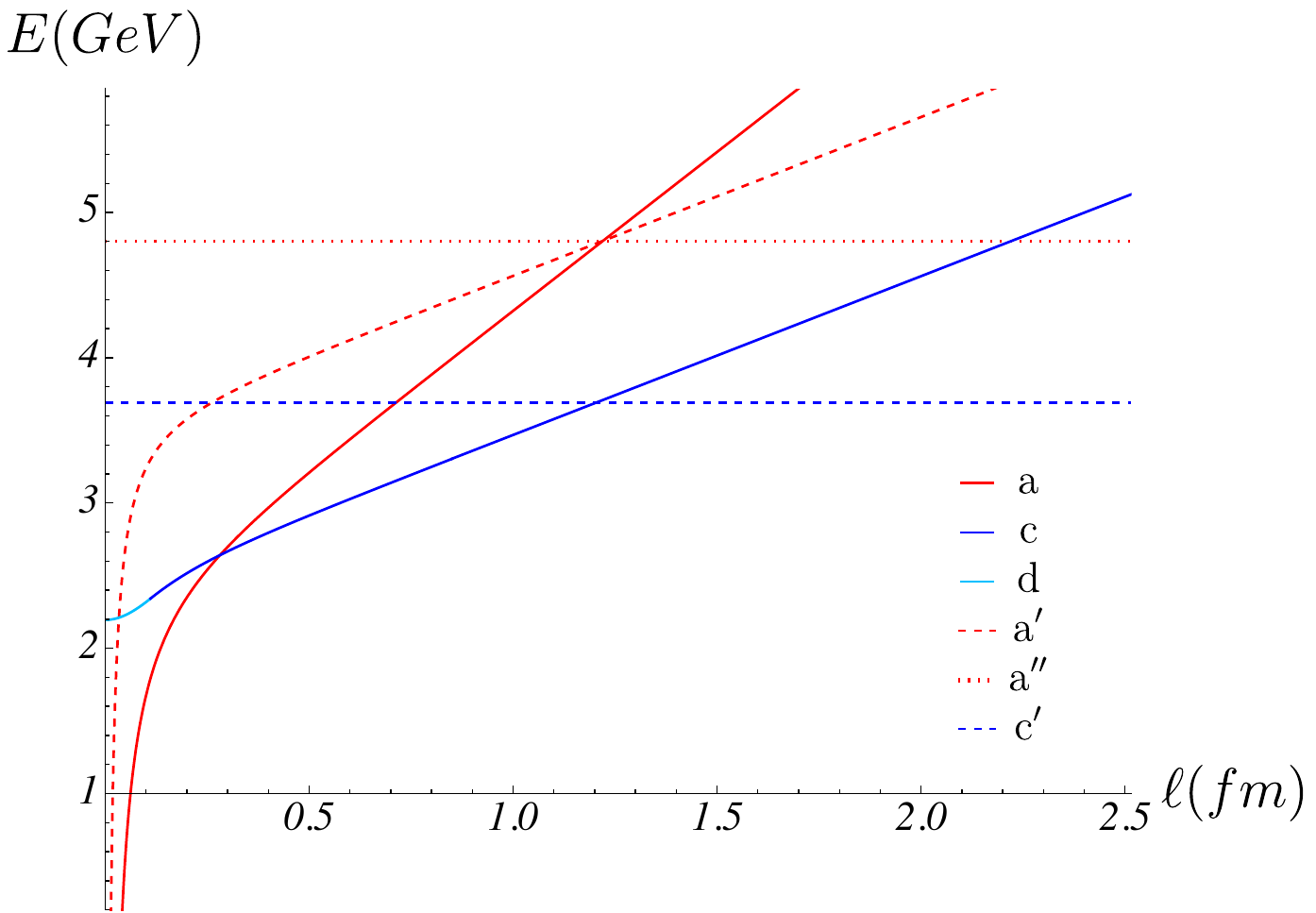}
\hspace{0.2cm}
\includegraphics[width=8.7cm]{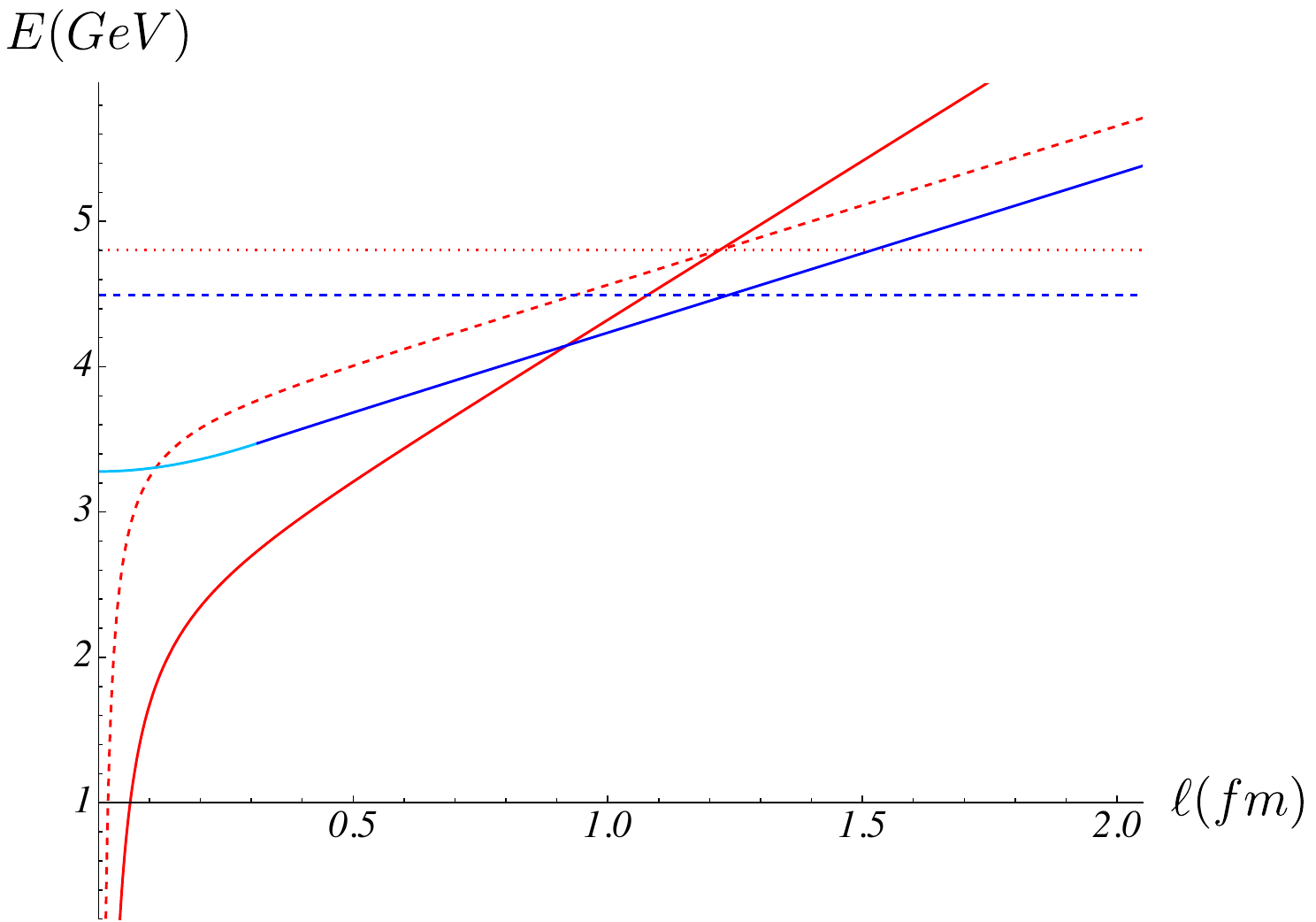}
\caption{{\small The energies of the string configurations for type-A ordering at $w=0.19\,\text{fm}$ (left) and at $w=0.54\,\text{fm}$ (right). }}
\label{E019054}
\end{figure}

Having understood the disconnected string configurations with light quarks, we can extend the analysis of section 4. For the limiting values of $w$, the energies of the relevant configurations are plotted in Fig.\ref{E019054}. We see that, in both   
\noindent cases, the ground  state potential is expressed in terms of three string configurations as $V_0=\min\{E^{\text{(a)}},E^{\text{(c)}},E^{\text{($\mathrm{c}'$)}}\}$. This suggests that the ground state is a mixed state of three states: two hadronic molecules and a tetraquark state associated with the tetraquark configuration (c). Its large-$\ell$ behavior is governed by the two-baryon configuration ($\mathrm{c}'$). The transition from configuration (c) to ($\mathrm{c}'$) can be interpreted as the breaking of the internal string through the creation of a light quark pair, as already discussed. This corresponds to the strong decay of the system into a pair of doubly heavy baryons: 

\begin{equation}\label{2baryon-decay}
QQ\bar Q\bar Q\to QQq+\bar Q\bar Q\bar q
\,.
\end{equation}

It is of physical interest to estimate the string breaking distance and, as a byproduct, one of the critical lengths defined in \eqref{lijc}. Following \cite{a-strb}, we define the length of string 3 as the distance between the projections of the baryon vertices onto the boundary (see Fig.\ref{tetra5d}). The string breaking distance is then $\boldsymbol{\ell}_{\4Q}=\vert Y\bar Y\vert$ evaluated at $\ell=\ell_c$, where $\ell_c$ is the critical length satisfying the equation $E^{\text{(c)}}(\ell)=E^{(\mathrm{c}')}(w)$ at fixed $w$. With the help of the expression leading to \eqref{EIl}, $\boldsymbol{\ell}_{\4Q}$ can be written explicitly as

\begin{equation}\label{l4Q}
	\boldsymbol{\ell}_{\4Q}=
	\frac{1}{\sqrt{\s}}{\cal L}(\bar\lambda,v)\Big\vert_{v=\vc}
	\,,
	\end{equation}
where $\vc$ is a solution of $E^{\,\i}(v)=2E_{\QQq}(w)$.\footnote{For the values of $w$ considered here, the large-$\ell$ behavior of $E^{\text{(c)}}$ is described by configuration (I).} Importantly, $\boldsymbol{\ell}_{\4Q}$ is finite and independent of the normalization constant $c$. Using this formula, we can make several estimates, which are summarized in Table \ref{estimates}. The conclusion then is that, in contrast to the string tension, the string breaking 
\begin{table*}[h]
\renewcommand{\arraystretch}{2}
\centering 	\scriptsize
\begin{tabular}{lccccccr}				
\hline
$w\,(\text{fm}) $ ~&~~ $E_{\QQq}\,(\text{GeV}) $  ~~&~~ $\ell_c\,(\text{fm})$ ~~~~&~~~~~ $\boldsymbol{\ell}_{\4Q}\,(\text{fm})$ ~&~~~~~ $\boldsymbol{\ell}_{\4Q}/\boldsymbol{\ell}_{\QQb}$
\rule[-3mm]{0mm}{8mm}
\\
\hline \hline
0.19  &    1.846  &  1.205  &~  1.137  &~~~  0.932 \\
0.31  &    1.990  &  1.194  &~  1.040  &~~~ 0.852 \\
0.37  &    2.057  &  1.200  &~  1.005  &~~~  0.824 \\
0.39  &    2.080  &  1.204  &~  0.995  &~~~  0.816 \\  
0.46  &    2.158  &  1.218  &~  0.964  &~~~  0.790 \\
0.54  &    2.246  &  1.236  &~  0.932  &~~~  0.764 \\
 \hline \hline
\end{tabular}
\caption{ \small Some estimates of the string breaking distance $\boldsymbol{\ell}_{\4Q}$. The values of $\ell_c$ are included to facilitate comparison with the lattice. Note that the function $\ell_c(w)$ has a local minimum near $w=0.29\,\text{fm}$. Here $\boldsymbol{\ell}_{\QQb}=1.22\,\text{fm}$.}
\label{estimates}
\end{table*}
distance is not universal. As in the $QQQ$ system \cite{astbr3Q}, it may be smaller than that in the $Q\bar Q$ system and, moreover, may depend on geometry, i.e. on the positions of the heavy quarks/antiquarks in space. As a further remark, we point out that the equation $E^{\text{(c)}}(\ell)=2E_{\QQq}(w)$ can be solved approximately. The reason for this is that in the vicinity of $\ell=1.22\,\text{fm}$ the function $E^{\text{(c)}}(\ell)$ is nearly linear, almost a straight line as seen from Fig.\ref{E019054}. If so, then, with the help of the asymptotic expansion \eqref{Ec-large}, we find 

\begin{equation}\label{lc}
\ell_c\approx 
\frac{1}{\sigma}\bigl(2 E_{\QQq}(w)-C^{\text{(c)}}\bigr)
\,.
\end{equation}
Here $C^{\text{(c)}}$ is given by the first expression in \eqref{Ec-large}. Importantly, $\ell_c$ is independent of the normalization constant $c$ as follows immediately from the expressions for $E_{\QQq}$ and $C^{\text{(c)}}$. It turns out that this is a very accurate approximation, which differs from the exact solution only by the 4 digit after the decimal point. 

The first excited B-O potential has a rather complicated structure. It is constructed from six string configurations and is written as $V_1=\min\{E^{\text{(a)}},E^{\text{(c)}},E^{\text{(d)}},E^{\text{($\mathrm{a}'$)}},E^{\text{($\mathrm{a}''$)}},E^{\text{($\mathrm{c}'$)}}\}$, as seen from Fig.\ref{E019054}. One of the new features, which may appear counterintuitive, is a modification of the potential at small quark separations due to string breaking. Indeed, configuration ($\mathrm{a}'$), obtained from configuration (a) by breaking one string, dominates at small $\ell$. At larger $\ell$, it transforms to configuration (d). This transition can be interpreted as a combination of $q\bar q$ pair annihilation, string intersecting, and $V\bar V$ pair creation at an intersection point. Such a process occurs at $\ell=0.034\,\text{fm}$ for $w=0.19\,\text{fm}$ and at $\ell=0.113\,\text{fm}$ for $w=0.54\,\text{fm}$.\footnote{Although the former length is smaller than a typical lattice spacing, the latter is not. This makes it possible that the effect of string breaking on $V_1$ at small quark separations could be seen on the lattice.} At intermediate $\ell$, the structure of $V_1$ remains the same as in section 4. At larger $\ell$, however, it expectedly changes owing to the disconnected configurations containing light quarks. First, a transition occurs from configuration (a) to configuration ($\mathrm{c}'$). For $w=0.19\,\text{fm}$ it occurs at $\ell=0.714\,\text{fm}$, while for $w=0.54\,\text{fm}$ at $\ell=1.078\,\text{fm}$. This transition can be interpreted as a combination of $V\bar V$ and $q\bar q$ pair creation. Then, a transition between configurations ($\mathrm{c}'$) and (c) occurs, as we have already seen for $V_0$. Finally, the potential gets flattened by a transition from configuration (c) to configuration ($\mathrm{a}''$). This transition occurs at $\ell=2.222\,\text{fm}$ for $w=0.19\,\text{fm}$ and at $\ell=1.521\,\text{fm}$ for $w=0.54\,\text{fm}$. It can be interpreted as a combination of annihilation of one $V\bar V$ pair and string breaking due to creation of two pairs of light quarks. 

\subsection{Type-B ordering}

The above analysis can be straightforwardly extended to type-B ordering. In this case the relevant string configurations are shown in Figs.\ref{c4B} and \ref{c4Bq}, which allow several simplifications. One immediate simplification is that we do not need to consider configurations (a), ($\mathrm{a}'$) and (d), as these have already been analyzed. Another simplification is that configuration (c) does not exist for B-type ordering \cite{a4Q}. We first work out the disconnected configurations (e) and ($\mathrm{e}'$), and then analyze the string configurations relevant for the two low-lying potentials.

\subsubsection{The mesonic configurations}

Consider first the two-meson configuration (e). Its energy is simply 

\begin{equation}\label{Ele}
	E^{\text{(e)}}=2E_{\QQb}(w)
	\,.
\end{equation}
Using the formulas of Appendix B, this can equivalently be written as  $E^{\text{(e)}}=4\g\sqrt{\s}\,{\cal E}^+(\lambda_{\text{e}})+4c$, where $\lambda_{\text{e}}$ is a solution of the equation $w=\frac{2}{\sqrt{\s}}{\cal L}^+(\lambda)$. 

Similarly, for the multimeson configuration ($\mathrm{e}'$), the energy is given by 

\begin{equation}\label{Elep}
	E^{\text{($\mathrm{e}'$)}}
	=
	E_{\QQb}(w)+2E_{\Qqb}
	\,.
\end{equation}
Thus, neither expression depends on $\ell$.

As a side remark, although configuration (c) does not exist for type-B ordering, configuration ($\mathrm{c}'$), which contains two doubly heavy baryons, does. However, it turns out that, in the range of widths considered in this section, it is irrelevant for the two low-lying B-O potentials. 

\subsubsection{More detail on the string configurations}

\begin{figure}[htpb]
\centering
\includegraphics[width=8.5cm]{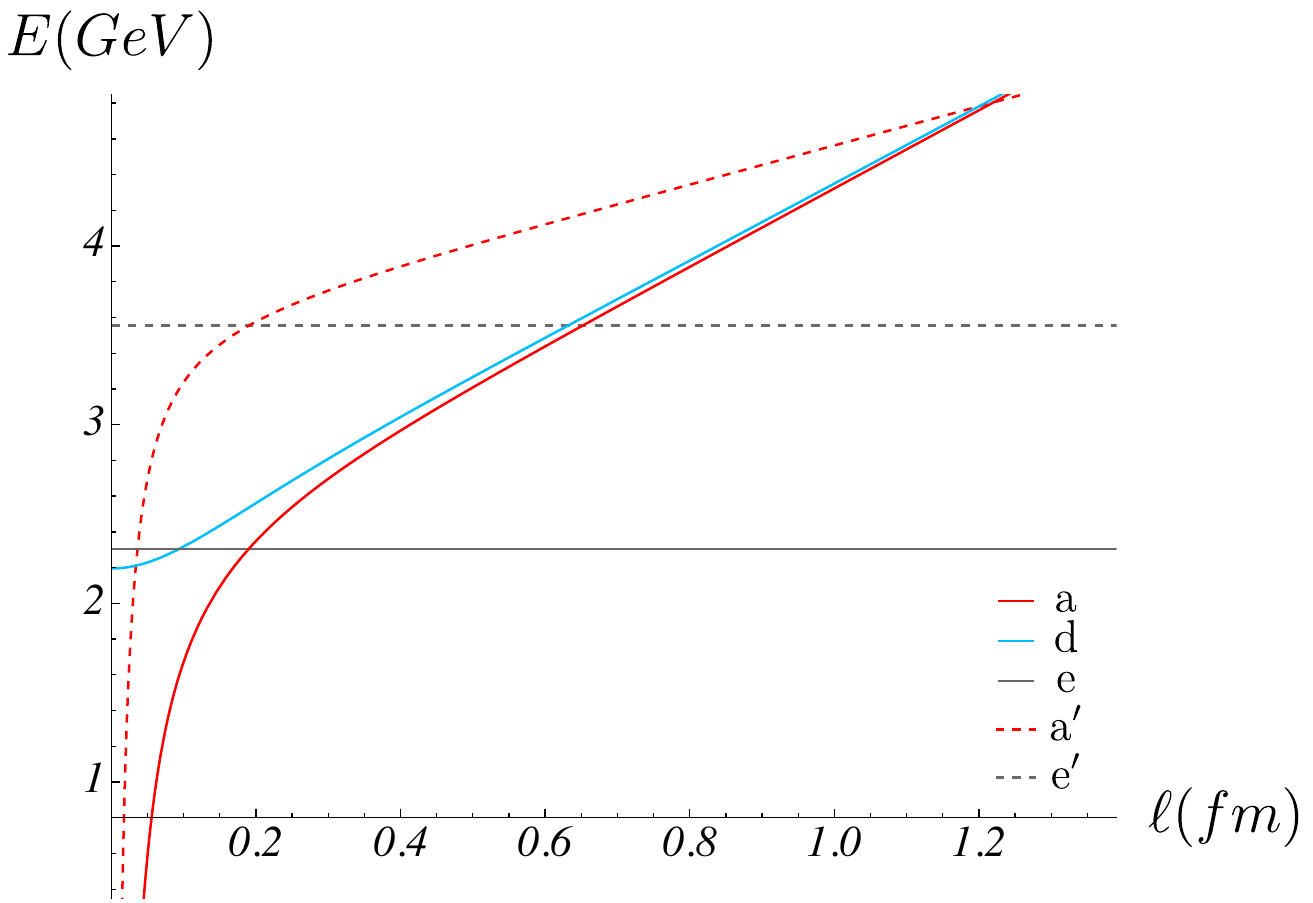}
\hspace{0.5cm}
\includegraphics[width=8.5cm]{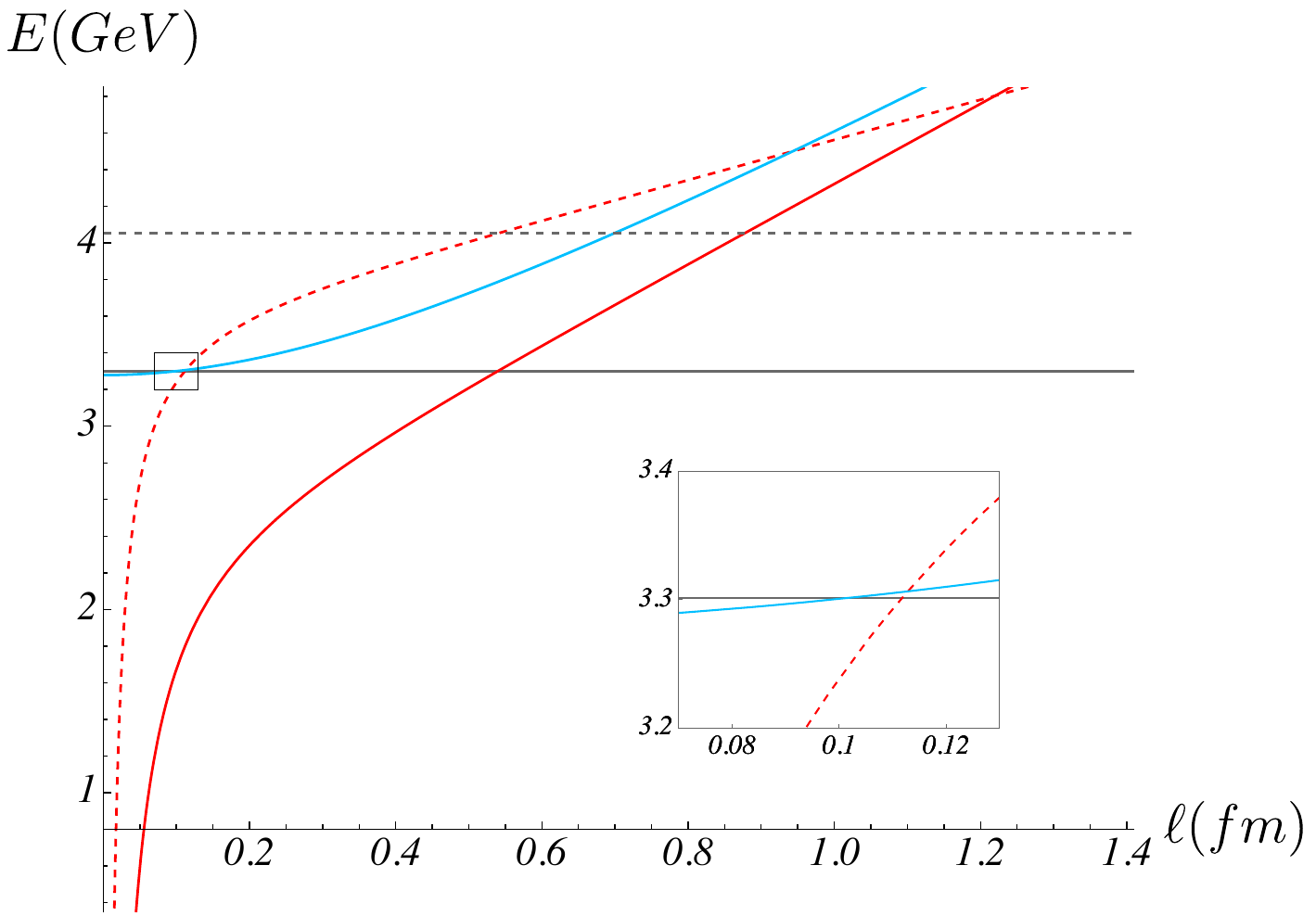}
\caption{{\small The energies of the string configurations for type-B ordering at $w=0.19\,\text{fm}$ (left) and at $w=0.54\,\text{fm}$ (right).}}
\label{EB19054}
\end{figure}

We are now ready to analyze the string configurations. In Fig.\ref{EB19054}, we plot their energies as functions of $\ell$ for the limiting values of $w$.
In both cases, the ground state potential is provided by the two-meson configurations (a) and (e), explicitly, $V_0=\min\{E^{\text{(a)}},E^{\text{(e)}}\}$.\footnote{In the literature, such a potential is called a flip-flop potential \cite{flip-flop}.}  The transition between them occurs at $\ell=0.190\,\text{fm}$ for $w=0.19\,\text{fm}$ and at $\ell=0.540\,\text{fm}$ for $w=0.54\,\text{fm}$. It corresponds to the decay of a hadronic molecule into two mesons
\begin{equation}\label{2meson-decay}
QQ\bar Q\bar Q\to 2Q\bar Q
\,,	
\end{equation}
and can be interpreted in terms of string interactions as string reconnection.  

The first excited potential again has a rather complicated structure. The string configurations contributing to it at the limiting values of $w$ are slightly different from each other, when carefully viewed together. That is because configuration (d), which is relevant at the lower bound, becomes irrelevant at the upper one. For $w=0.19\,\text{fm}$, the potential is given by $V_1=\min\{E^{(\mathrm{a})},E^{(\mathrm{a}')},E^{\text{(d)}},E^{\text{(e)}},E^{(\mathrm{e}')}\}$ so that configuration ($\mathrm{a}'$) dominates at very small $\ell$. A transition to configuration (d) occurs at $\ell=0.034\,\text{fm}$ and can be interpreted as a combination of the annihilation of a $q\bar q$ pair, the creation of a $V\bar V$ pair, and string reconnection. A further transition occurs at $\ell=0.094\,\text{fm}$, due to the annihilation of the $V\bar V$ pair. It brings configuration (d) to configuration (e). By contrast, at $w=0.54\,\text{fm}$ a transition from configuration ($\mathrm{a}'$) goes directly to configuration (e), bypassing configuration (d). This occurs at $\ell=0.112\,\text{fm}$. In this case, the first excited potential is given by $V_1=\min\{E^{\text{(a)}},E^{(\mathrm{a}')},E^{\text{(e)}},E^{(\mathrm{e}')}\}$. Then, for both configurations, a transition between configurations (e) and (a) occurs, as discussed above for $V_0$. Finally, the potential becomes flattened owing to a transition to configuration ($\mathrm{e}'$). This transition occurs at rather large $\ell$, namely at $\ell=0.652\,\text{fm}$ for the lower bound and at $\ell=0.877\,\text{fm}$ for the upper bound. It can be interpreted as a combination of the creation of a $q\bar q$ pair and string reconnection. 

\section{The potentials}
\renewcommand{\theequation}{4.\arabic{equation}}
\setcounter{equation}{0}
\subsection{Some details about the potentials}

Once the string configurations are understood, the diagonal elements of the model Hamiltonian \eqref{H} are known. It remains challenging, however, to compute the off-diagonal elements within the effective string model. In principle, these elements can be determined from a correlation matrix in lattice QCD. Therefore, it is impossible to visualize the exact form of the potentials. Nevertheless, we can gain valuable insight from our experiences with other quark systems, particularly by using the approximate magnitudes of the $\Theta$ values near the transition points.\footnote{For instance, one may assume that the $\Theta$'s are approximately constant, with values around $47\,\text{MeV}$, as found for the $Q\bar Q$ system on the lattice \cite{bulava}.} 

\begin{figure}[htpb]
\centering
\includegraphics[width=8.5cm]{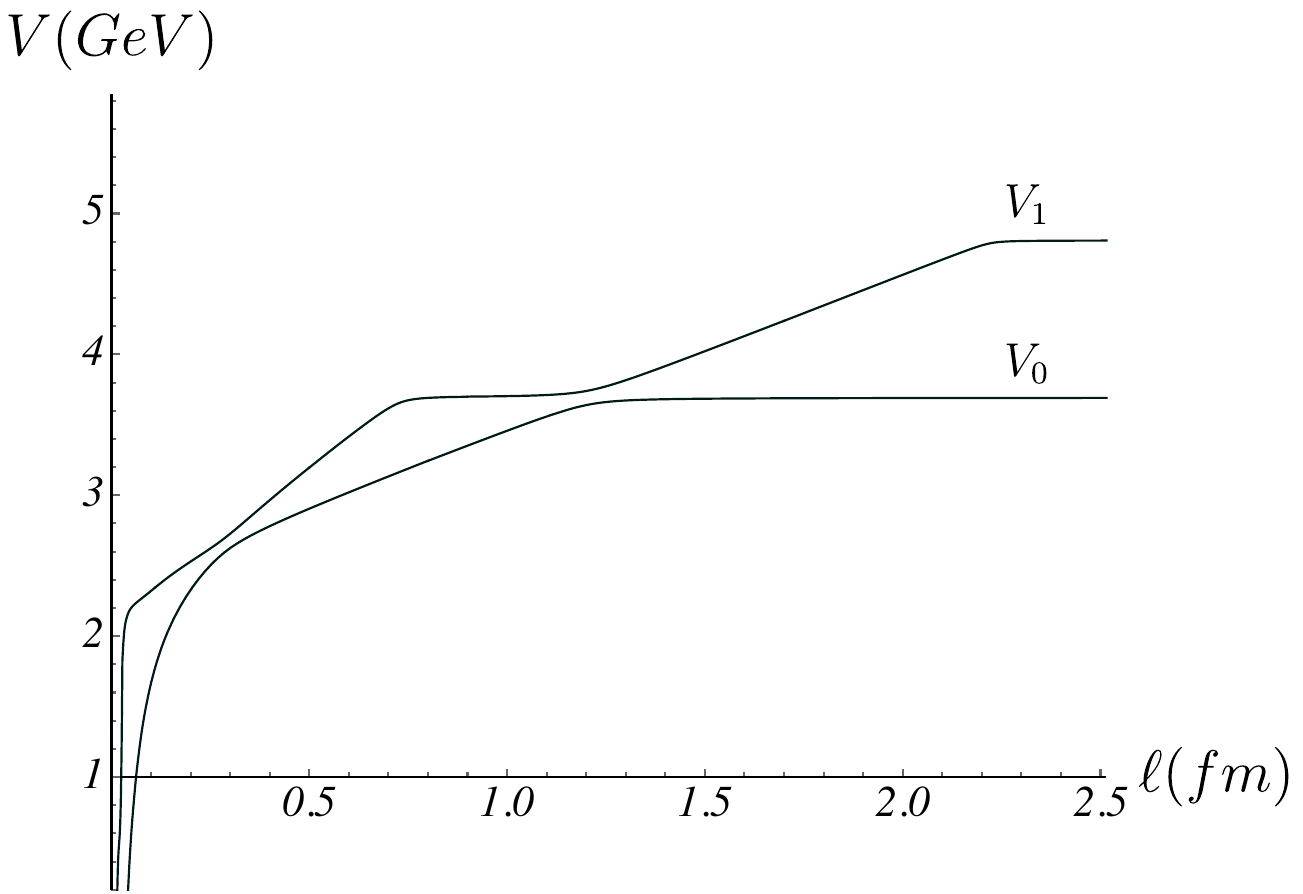}
\hspace{0.5cm}
\includegraphics[width=8.5cm]{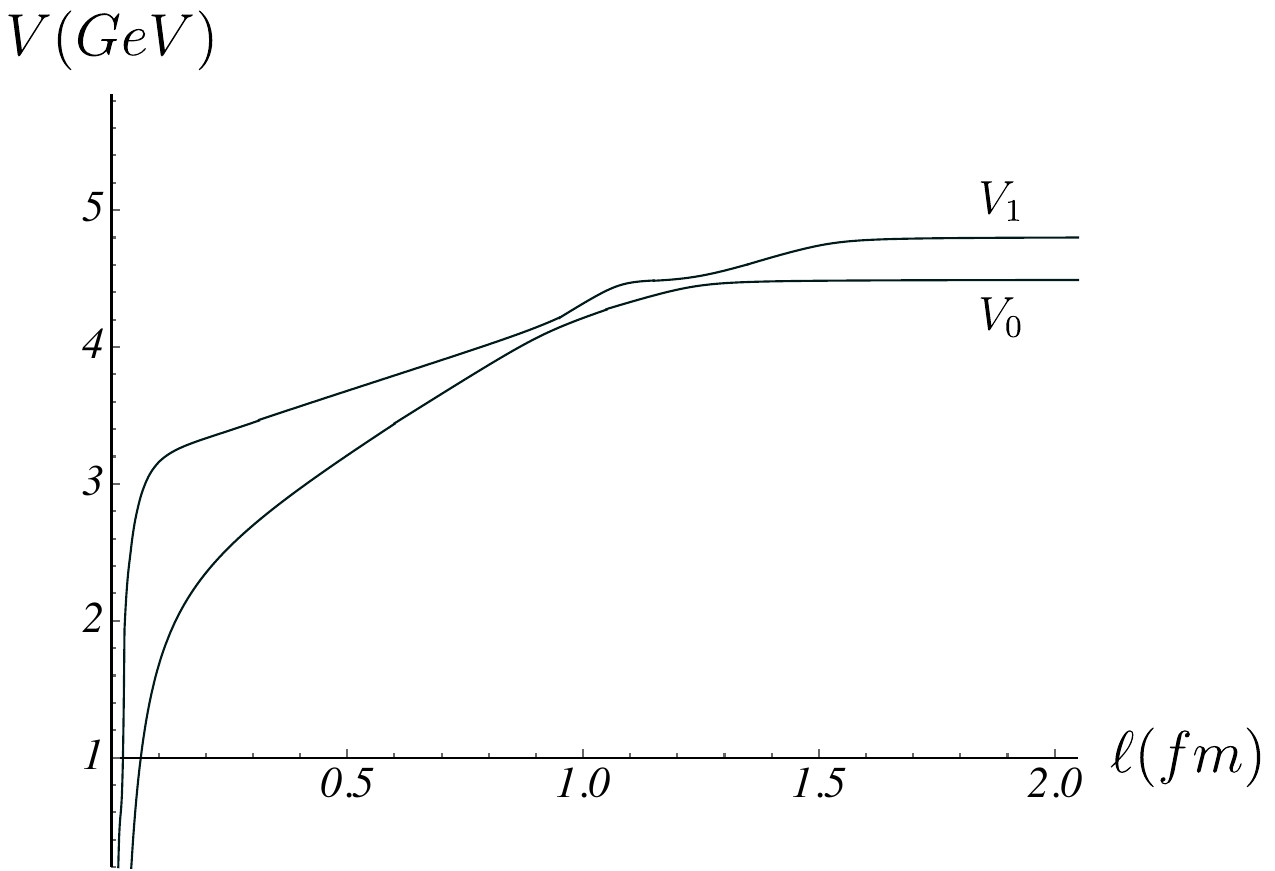}
\caption{{\small Sketched here are the two lowest B-O potentials for type-A ordering at $w=0.19\,\text{fm}$ (left) and at $w=0.54\,\text{fm}$ (right). The formula \eqref{V0QQb} helps with plotting the potentials in the vicinity of transition points.}}
\label{V0V1}
\end{figure}

For type-A ordering, this leads us to the overall picture sketched in Fig.\ref{V0V1}. Although it includes only the results for the limiting values
of $w$, the same picture holds for intermediate values. Two conclusions can be drawn from these results. The first is that at large $\ell$ the gap between $V_0$ and $V_1$ decreases with $w$. This is because the energy $E^{(\mathrm{c}')}$ increases with $w$, while $E^{(\mathrm{a}'')}$ remains constant. The second is that for $V_0$ the support of the tetraquark configuration (c) shrinks from $0.282\,\text{fm}\leq \ell\leq 1.205\,\text{fm}$ at $w=0.19\,\text{fm}$ to $0.920\,\text{fm}\leq \ell\leq 1.236\,\text{fm}$ at $w=0.54\,\text{fm}$. As we will see in Sec.V, both conclusions are related to a qualitative change in the structure of the potentials at larger widths.  

\begin{figure}[htpb]
\centering
\includegraphics[width=8.30cm]{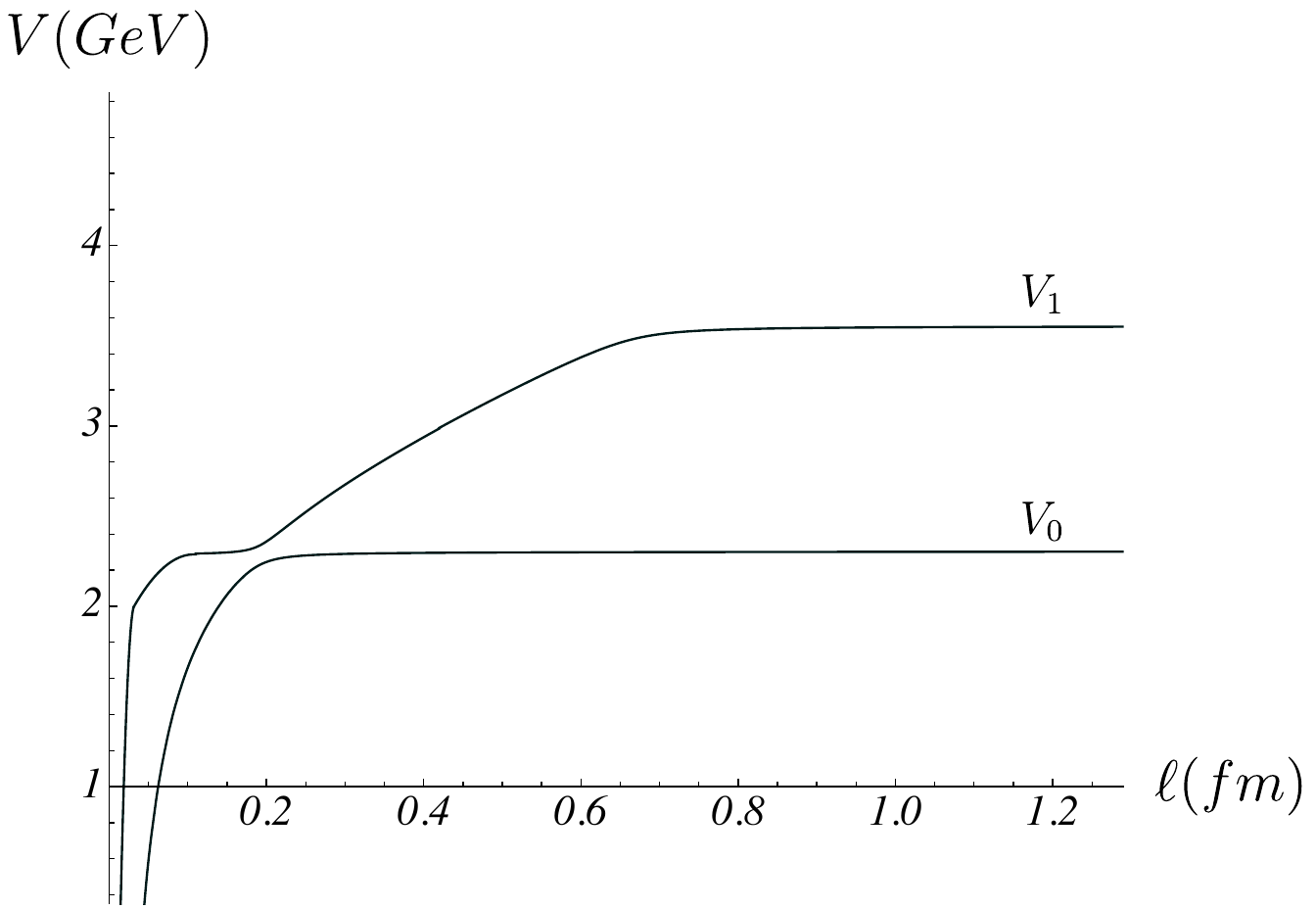}
\hspace{1cm}
\includegraphics[width=8.15cm]{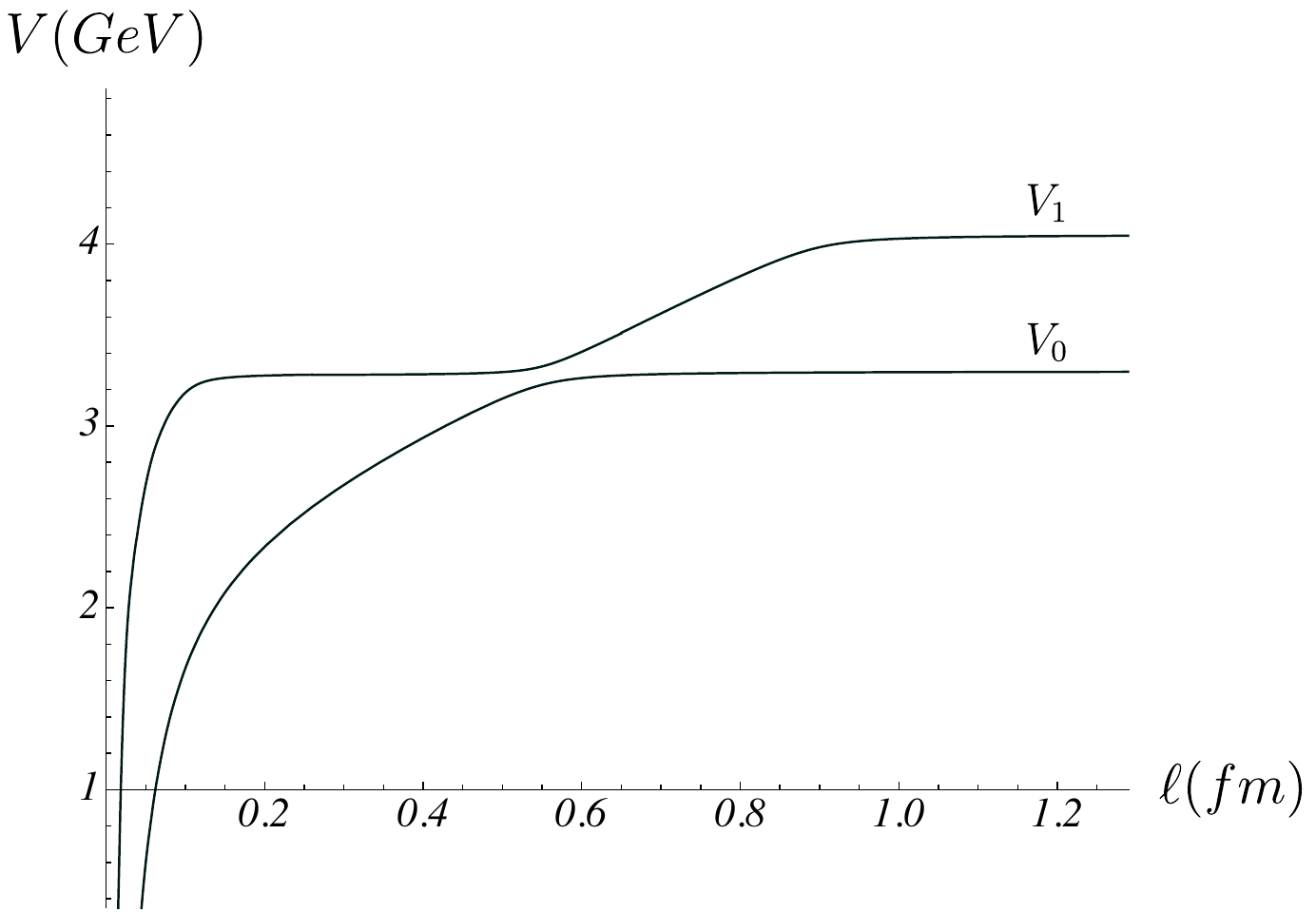}
\caption{{\small Sketched here are the two lowest B-O potentials for B-type ordering at $w=0.19\,\text{fm}$ (left) and at $w=0.54\,\text{fm}$ (right).}}
\label{BV0V1}
\end{figure}

Similarly, one can gain insight into the potentials for B-type ordering. The resulting plots are presented in Fig.\ref{BV0V1}. 
\noindent In this case, the gap between $V_0$ and $V_1$ also decreases with $w$ at large $\ell$. The reason is that the energy $E^{(\mathrm{e})}$ increases with $w$ faster than $E^{(\mathrm{e}')}$. Another observation is that, for $V_0$, the support of configuration (a) increases with $w$. Therefore, we expect the structure of $V_0$ and $V_1$ to change qualitatively at larger $w$. We return to this issue in the next section.

\subsection{Comparison with lattice QCD}

For the $QQ\bar Q\bar Q$ system, the results of numerical simulations are rather selective and are limited to theories with $N_f=0$ and $N_f=2+1$ flavors \cite{alexa,suga, bicu4Q}. In the latter case, the effects of string breaking have not been considered. Nevertheless, it is worthwhile to compare our results qualitatively with those obtained in lattice QCD. We begin with type-A ordering.

In the quenched case, the ground state potential $V_0$ was first studied in \cite{alexa,suga}. The main conclusion is that $V_0$ can be described in terms of the two-meson configuration (a) and the tetraquark configuration (c), with the former dominating at small $\ell$ and the latter at larger $\ell$, where it yields the double $Y$-law. Later, it was re-examined in \cite{bicu4Q}, with a conclusion that the large-$\ell$ behavior is better described by a triplet-antitriplet potential, which is a sum of the quark-antiquark potentials.\footnote{Except for the potentials associated with the two-meson configurations (a) and (e), the other potentials used in the fits in \cite{bicu4Q} have no string description.} This may remind the reader of the debates concerning the $Y$- and $\Delta$-laws in the IR limit of the $QQQ$ system: the $Y$-law corresponds to a genuine three body interaction, whereas the $\Delta$-law is a sum of two body interactions. The issue was ultimately resolved in favor of the $Y$-law (see, e.g., \cite{pdf} and references therein). As in the $QQQ$ case, where our model yields the $Y$-law \cite{a3Q}, the results of Sec.III support the original conclusion of \cite{alexa,suga}, namely the double $Y$-law. Much less is known about the first excited potential $V_1$. Although it was studied in \cite{bicu4Q}, it is unclear what conclusions can be drawn from the fits. On the other hand, our results suggest that it can be described in terms of the two-meson, tetraquark, and pinched tetraquark configurations (see Fig.\ref{E01054s}). 

In the presence of light quarks, both potentials were studied in \cite{bicu4Q}, but only for $\ell$ up to $0.91\,\text{fm}$, with conclusions similar to those in the quenched case. Our results, however, suggest that, for sufficiently large $\ell$ but smaller than $1\,\text{fm}$, the ground state potential is described in terms of the tetraquark configuration which leads to the double $Y$-law rather than to the triplet-antitriplet potential. More precisely, this is true if $w$ is near its lower bound (see Fig.\ref{V0V1}). Nevertheless, we find agreement with \cite{bicu4Q}: the range of $\ell$ in which the two-meson configuration dominates $V_0$ increases with $w$. Of course, the potentials are modified by including the disconnected string configurations. Their contributions change not only the large-$\ell$ behavior but also the small-$\ell$ behavior (see Fig.\ref{E019054}). In particular, the two-baryon configuration $(\text{c}')$ makes the potential $V_0$ flatten if $\ell$ exceeds $1.19\,\text{fm}$. Another significant change is that, at small $\ell$,  the multimeson configuration ($\text{a}'$) dominates $V_1$. This occurs at rather small lengths, $\ell \lesssim 0.11\,\text{fm}$, and is therefore not visible in \cite{bicu4Q}. Hopefully, these predictions of our model can be verified in future numerical simulations. 

For type-B ordering, both potentials were studied only in \cite{bicu4Q}. There, it was concluded that, independently of the flavor content, the ground state potential can be described by the flip-flop potential, which arises from a superposition of the two-meson configurations (a) and (e). This conclusion agrees with the results in Sec.III. However, the first excited potential remains puzzling. Our results suggest that, in the quenched case, $V_1$ can be described in terms of the pinched tetraquark configuration and the mesonic configurations. In the presence of light quarks, the picture is modified by the multimeson configurations (see Fig.\ref{EB19054}). They modify both the large- and small-$\ell$ behaviors of $V_1$. These are our predictions for the first excited potential for type-B ordering.

\section{Further comments}
\renewcommand{\theequation}{5.\arabic{equation}}
\setcounter{equation}{0}
\subsection{Other excitations}

\begin{figure}[htbp]
\centering
\includegraphics[width=9cm]{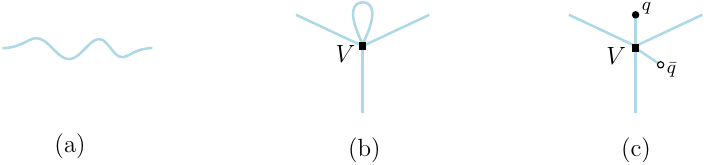}
\caption{{\small Some elementary excitations: (a) an excited string and (b)-(c) higher baryon vertices. Excited strings are denoted by wavy lines, here and below. Shown in (c) are sea quarks.}}
\label{excitations}
\end{figure}

The assumptions made in Sect.II regarding the elementary excitations are oversimplified for various reasons. In particular, excited strings such as the one sketched in Fig.\ref{excitations}a should be included when constructing string configurations 
for higher B-O potentials of multiquark systems. These strings represent a class of gluonic excitations that have been extensively studied in lattice QCD, but primarily in the context of the $Q\bar Q$ system. Similarly, excitations arising from the baryon vertex, regarded as a five-brane, must also be taken into account. This leads to a tower of higher (excited) vertices.\footnote{If so, then the string junction in four dimensions should be viewed as an object that may be in multiple excited states. It would therefore be highly desirable to seek evidence for such excitations through lattice simulations.} The simplest examples of those are shown in Figs.\ref{excitations}b-c, where the excitations originate from open strings terminating both on the brane and on light sea quarks. 

\begin{figure}[htbp]
\centering
\includegraphics[width=11cm]{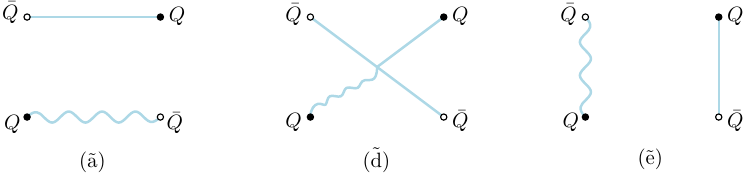}
\caption{{\small Four-dimensional string configurations including excited strings for type-B ordering. They are labeled by the same letters as the parent configurations in Fig.\ref{c4B}, from which they can be obtained by replacing an unexcited string with an excited one.}}
\label{c4Bhybrid}
\end{figure}

We will illustrate this with one simple example. As in \cite{a3Q}, we restrict ourselves to one type of string excitations, namely the type related to the hybrid potential $\Sigma_u^-$ \cite{kuti}. It can be modeled within the five-dimensional string model we are pursuing \cite{ahybrid}. For simplicity, we further restrict to type-B ordering,  for which the string configurations are shown in Fig.\ref{c4B}. An extension in which one excited string is included in each configuration is straightforward. The corresponding configurations are shown in Fig.\ref{c4Bhybrid}. 
One immediate simplification is that configuration ($\tilde{\mathrm{d}}$) contributes to neither the potential $V_0$ nor the potential $V_1$.\footnote{If one replaces a baryon vertex in configuration (d) with any higher vertex, as those shown in Fig.\ref{excitations}b-c, one can make a similar argument.} This follows from the fact that configuration (d) can contribute only to $V_1$. Therefore, it suffices to consider configurations $(\tilde{\mathrm{a}})$ and $(\tilde{\mathrm{e}})$ only. These can be treated in a way similar to the parent configurations ($\mathrm{a}$) and ($\mathrm{e}$). 

For configuration ($\tilde{\mathrm{a}}$), the energy is simply 

\begin{equation}\label{ahybrid}
	E^{\text{($\tilde{\mathrm{a}}$)}}(\ell)
	=E_{\QQb}(\ell)
	+
	\tilde{E}_{\QQb}(\ell) 
	\,,
\end{equation}
with the function $\tilde{E}_{\QQb}$ defined parametrically in \eqref{EQQb-hybrid}. Since, at very small quark separations, the behavior of $\tilde{E}_{\QQb}$ deviates from what is expected from perturbative QCD (see Appendix B), it is meaningless to discuss the small-$\ell$ expansion. By contrast, the large-$\ell$ expansion is meaningful and it reads 

\begin{equation}\label{ahybrid-large}
E^{\text{($\tilde{\mathrm{a}}$)}}(\ell)
=
2\sigma\ell+C^{(\tilde{\mathrm{a}})}+o(1)
\,,\qquad\text{with}\qquad 
C^{(\tilde{\mathrm{a}})}
=C_{\QQb}
+
\tilde{C}_{\QQb}
\,,
\end{equation}
where $\tilde{C}_{\QQb}$ is given by \eqref{EQQb-hybrid-large}. 

Similarly, for configuration ($\tilde{\mathrm{e}}$), the energy can be written as  

\begin{equation}\label{ehybrid}
	E^{\text{($\tilde{\mathrm{e}}$)}}
	=E_{\QQb}(w)
	+
	\tilde{E}_{\QQb}(w) 
	\,,
\end{equation}
which is a constant independent of $\ell$. 

\begin{figure}[htbp]
\centering
\includegraphics[width=8.6cm]{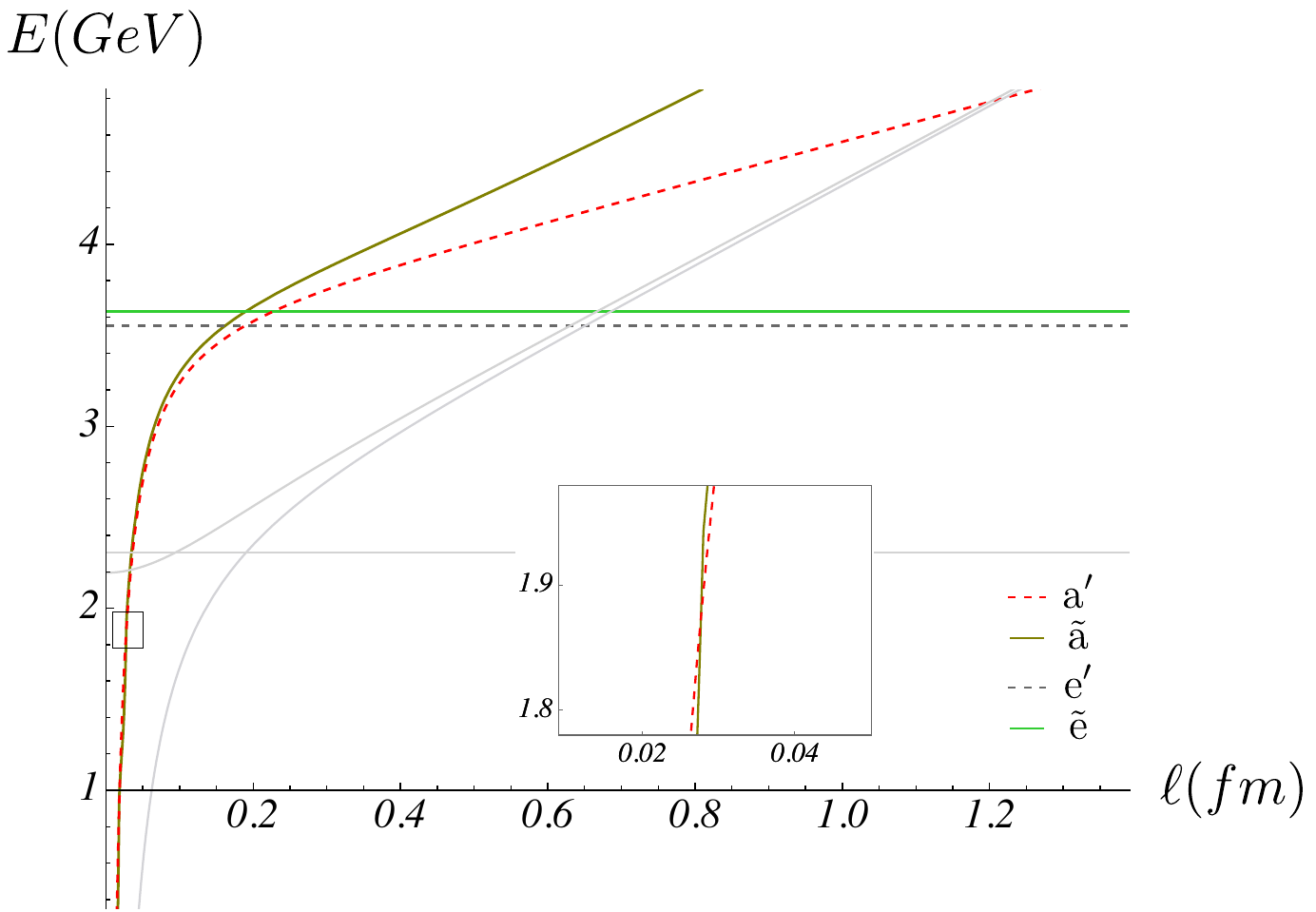}
\hspace{0.4cm}
\includegraphics[width=8.6cm]{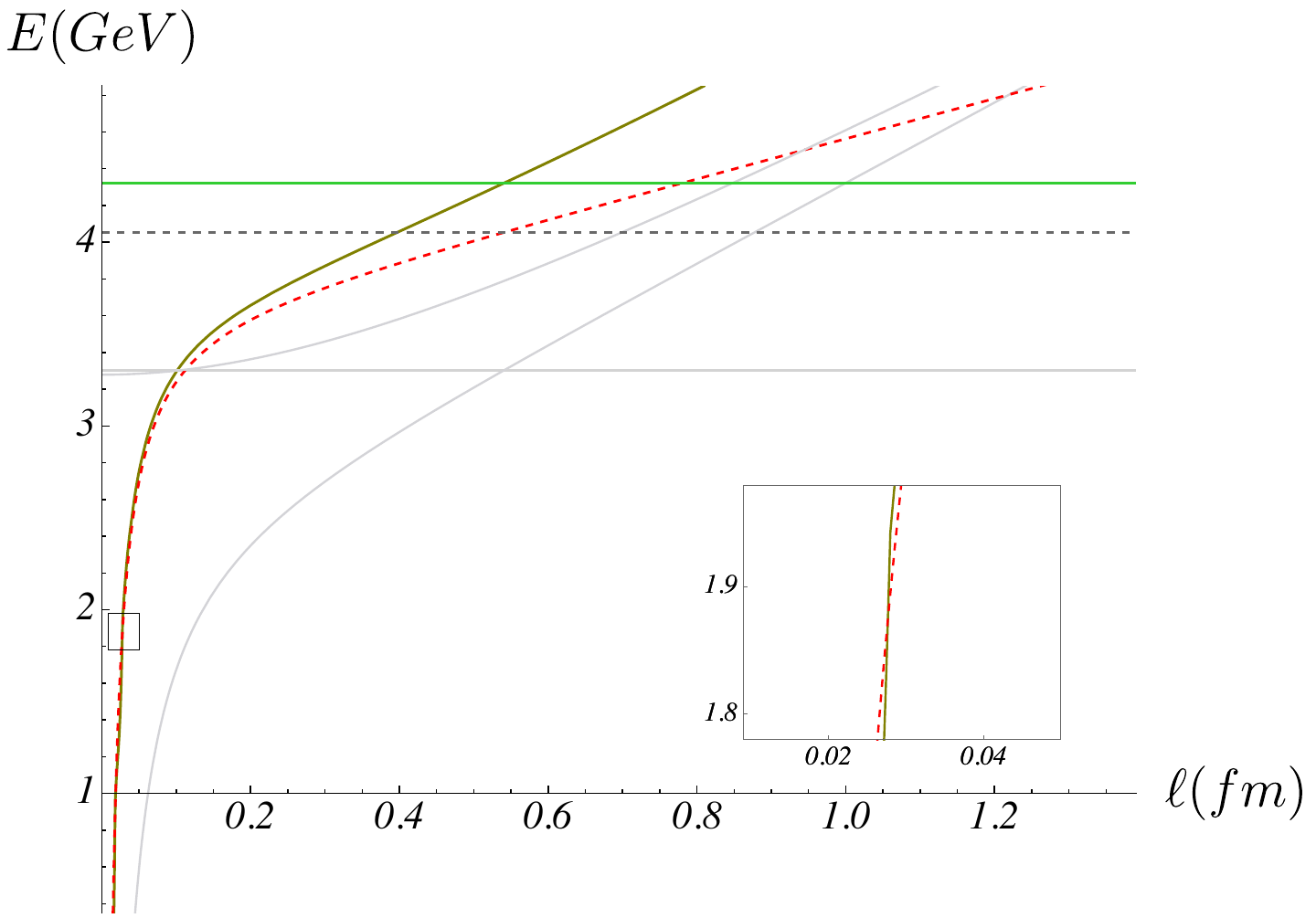}
\caption{{The plots of Fig.\ref{EB19054}, supplemented by the plots for configurations $(\tilde{\mathrm{a}})$ and $(\tilde{\mathrm{e}})$, at $w=0.19\,\text{fm}$ (left) and at $w=0.54\,\text{fm}$ (right). The plots irrelevant to our discussion are shown in gray.}}
\label{BHE019054}
\end{figure}

Now that we have computed the energies of the configurations with excited strings, a natural question arises as to whether they contribute to the potential $V_1$. We will be rather brief with this question, as it can not be resolved definitively. In Fig.\ref{BHE019054}, we plot the energies as functions of $\ell$ for the two limiting values of $w$. Two important points are worth noting. First, 
a simple estimate shows that the gap between $E^{(\mathrm{e}')}$ and $E^{(\tilde{\mathrm{e}})}$ is approximately $77\,\text{MeV}$ for $w=0.19\,\text{fm}$ and $268\,\text{MeV}$ for $w=0.54\,\text{fm}$. These values are compatible with the difference between the $\Sigma_u^-$ and $\Pi_u$ hybrid potentials, regarded as functions of $w$, which is approximately $77\,\text{MeV}$ and $218\,\text{MeV}$, respectively \cite{kuti}.\footnote{Note that $\Pi_u$ is the hybrid potential at level $N=1$, whereas $\Sigma_u^-$ is at level $3$.} This suggests that, in principle, the string excitations associated with $\Pi_u$ may be relevant to $V_1$. Second, at very small separations, $E^{(\tilde{\mathrm{a}})}$ dominates $V_1$. This is a subtle point, and it should be noted that for $\tilde{E}_{\QQb}$ we do not consider separations smaller than the lattice spacing (see Appendix B on more on this). 

We can summarize all of this by saying that achieving a clear understanding of the first excited potential remains a significant challenge.

\subsection{Larger $w$}

We have assumed the width in the range $0.19\,\text{fm}\leq w\leq 0.54\,\text{fm}$, which is appropriate for ground state diquarks. To gain deeper insight, it is important to consider excited diquarks, which are larger in size. We therefore extend our previous analysis to larger widths to partially address this issue. To avoid subtleties related to the excited potential $V_1$, we restrict our discussion to the ground state potential $V_0$.

Let us begin with type-A ordering. As discussed in subsection 4 of Sec.IIIA, $V_0$ is described in terms of configurations (a) and (c). In the absence of light quarks, this conclusion remains valid for arbitrary $w$ \cite{a4Q}. What happens to the potential in the presence of light quarks? A partial answer, given in subsection 6, is that, in the range under consideration, the potential is flattened by string breaking which occurs in configuration (c). Physically this means that a tetraquark state decays into two doubly heavy baryons (see \eqref{2baryon-decay}). Consequently, the potential is given by $V_0=\min\{E^{(\mathrm{a})},E^{(\mathrm{c})},E^{(\mathrm{c'})}\}$. There is, however, one simple but important observation that should be added to this conclusion. The energies of configurations (c) and $(\mathrm{c}')$ increase with $w$, whereas the energy of configuration (a) does not, as also seen in Fig.\ref{E019054}. An immediate consequence is that configuration ($\mathrm{a}''$) must be included in the analysis, since its energy being independent of $w$ becomes smaller than $E^{(\mathrm{c'})}$ at sufficiently large values of $w$. Further analysis shows that 

 \begin{equation}\label{V0w}
V_0=
\left\{ \begin{array}{lll} 
\min\{E^{(\mathrm{a})},E^{(\mathrm{c})},E^{(\mathrm{c'})}\}&, 
& w\leq 0.681\,\text{fm}\,,\\
\min\{E^{(\mathrm{a})},E^{(\mathrm{a''})},E^{(\mathrm{c})}\}&,
& 0.681\,\text{fm} \leq w \leq 0.711\,\text{fm}\,,\\
\min\{E^{(\mathrm{a})},E^{(\mathrm{a''})}\}&,
& w\geq 0.711\,\text{fm}\,.
\end{array} \right.
\end{equation}
Here $w=0.681\,\text{fm}$ is a solution of the equation $E^{(\mathrm{c'})}(w)=E^{(\mathrm{a''})}$, whereas $w=0.711\,\text{fm}$ is a solution of the equation $E^{(\mathrm{c})}(\ell_{\QQb},w)=E^{(\mathrm{a''})}$, where $\ell_{\QQb}$ is the string breaking distance for the $Q\bar Q$ system, defined in \eqref{LcQQb}. Thus, for larger $w$, the potential is flattened by double string breaking which occurs in configuration (a) and leads to the decay 

\begin{equation}\label{4meson-decay}
QQ\bar Q\bar Q\to 2Q\bar q+2q\bar Q
\,.
\end{equation}

\begin{figure}[htbp]
\centering
\includegraphics[width=8.6cm]{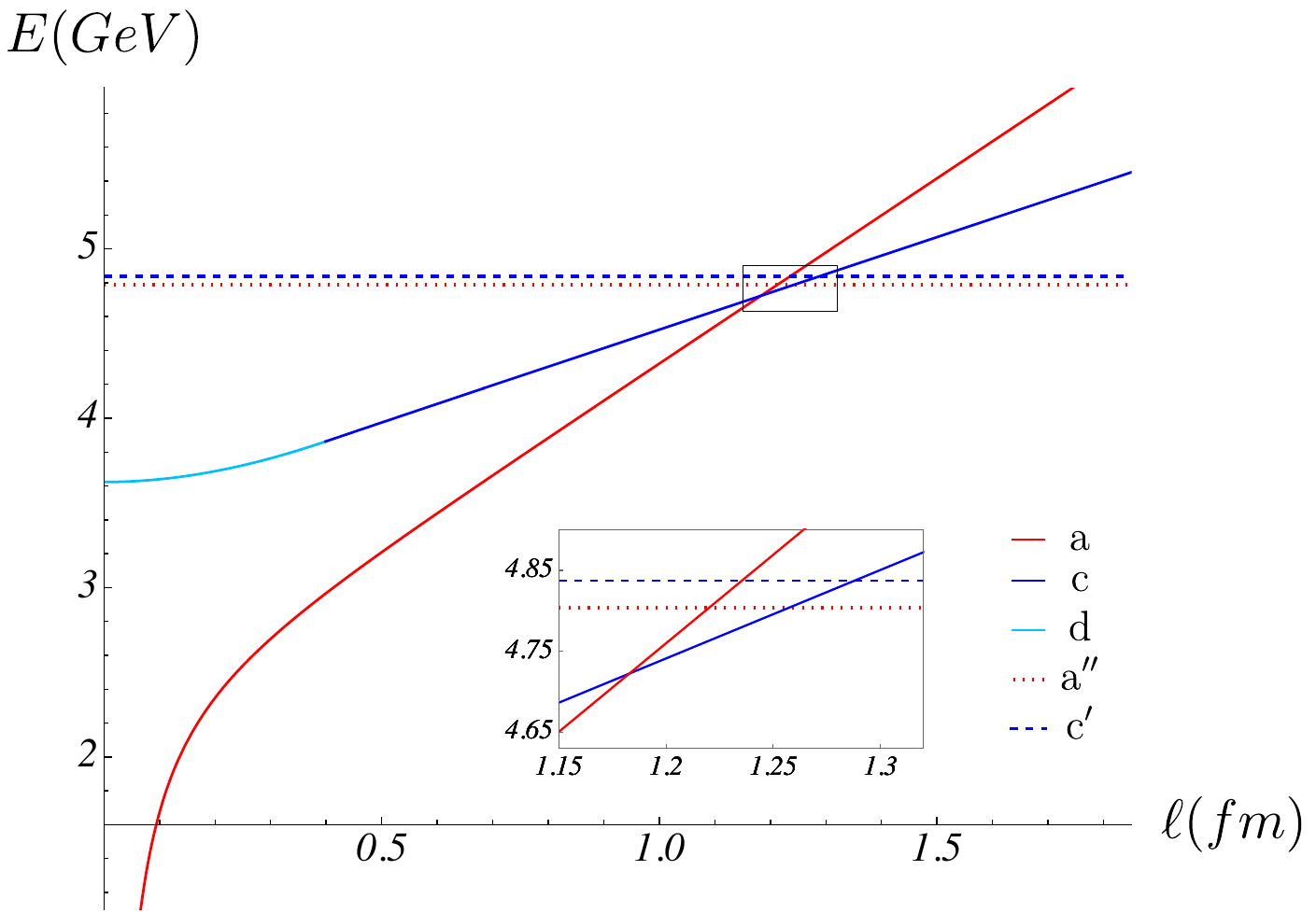}
\hspace{0.4cm}
\includegraphics[width=8.5cm]{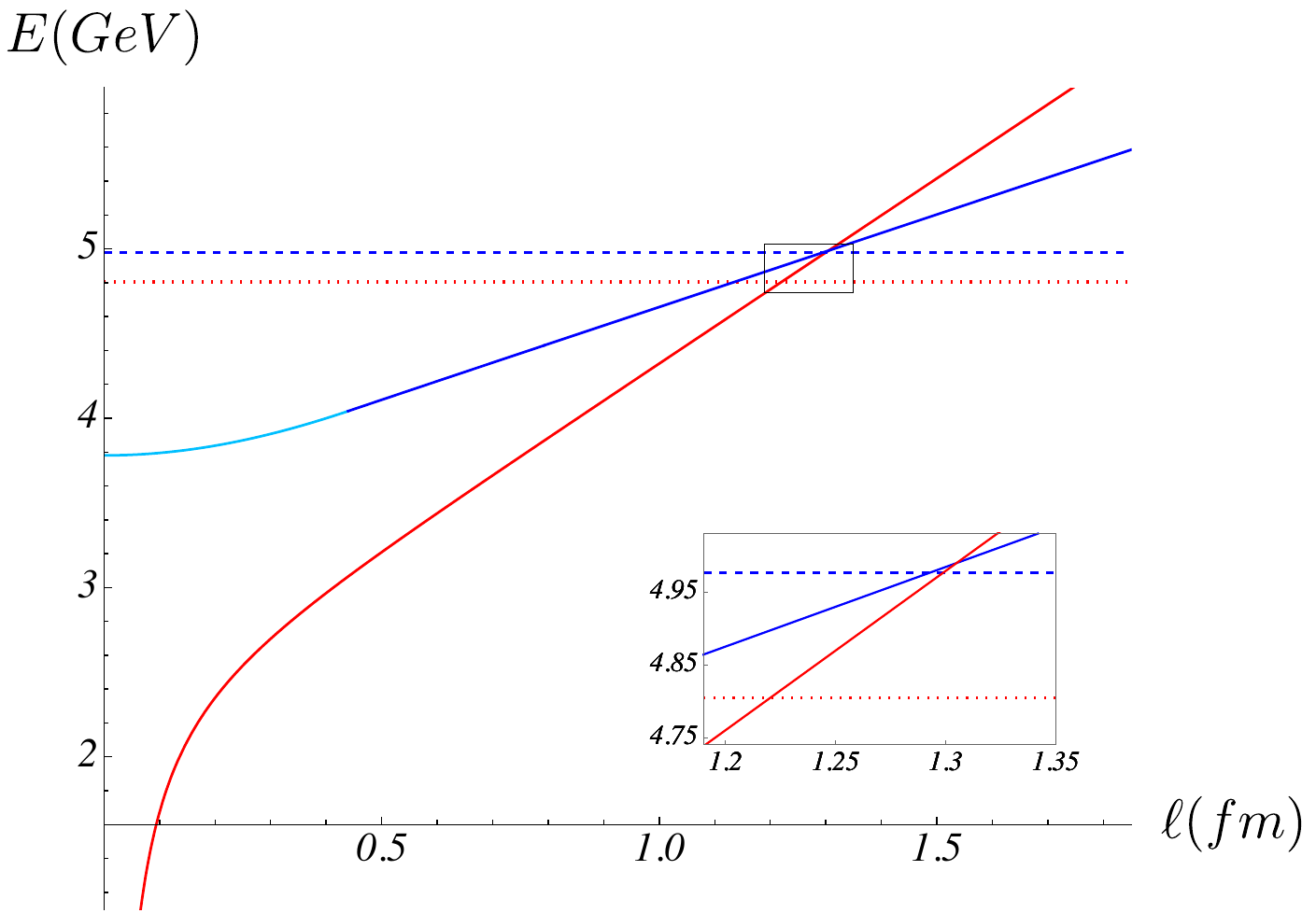}
\caption{{\small The energies of the string configurations included in \eqref{V0w} at $w=0.69\,\text{fm}$ (left) and at $w=0.76\,\text{fm}$ (right).}}
\label{E069-076}
\end{figure}

We conclude our discussion of type-A ordering with simple examples of the energies of the string configurations in \eqref{V0w} for the last two cases. These examples are shown in Fig.\ref{E069-076}. 

To perform a similar analysis for type-B ordering, we need to slightly change our terminology. Configurations ($\mathrm{a}'$) and ($\mathrm{e}'$) are obtained from configurations (a) and (e), respectively, by breaking either of the two strings, as shown in Fig.\ref{c4Bq}. Further string breaking leads, in both cases, to the same multimeson configuration, which we have already denoted by ($\mathrm{a}''$). In this subsection we denote it by ($\mathrm{e}''$) solely to keep track of its origin. 

\begin{figure}[htbp]
\centering
\includegraphics[width=7.75cm]{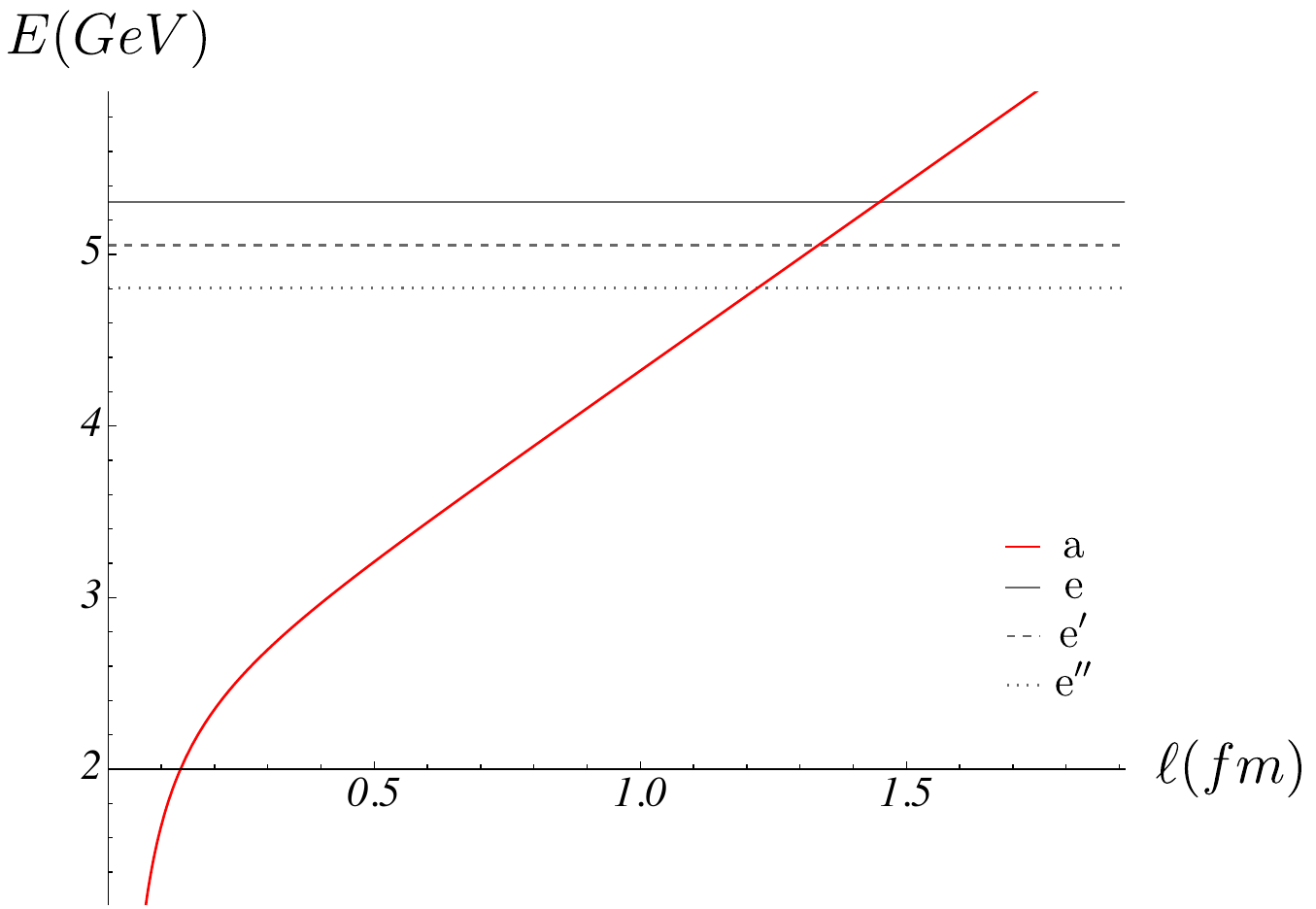}
\caption{{\small The energies of the string configurations included in \eqref{V0wB} at $w=1.45\,\text{fm}$. We include $E^{(\mathrm{e}')}$ for completeness. 
}}
\label{EB145}
\end{figure}

As we have seen in Sec.III, in the range of $w$ considered here, the ground state potential is given by $V_0=\min\{E^{(\mathrm{a})},E^{(\mathrm{e})}\}$. A key observation is that the energy of the disconnected configuration (e) increases with $w$, whereas that of $\mathrm{e}''$ remains constant. This implies that $E^{(\mathrm{e})}(w)>E^{(\mathrm{e}'')}$ for sufficiently large values of $w$. Because of this, configuration ($\mathrm{e}''$) must be included in the analysis. It is easy to see that the potential can be written as 

\begin{equation}\label{V0wB}
V_0=
\left\{ \begin{array}{lll} 
\min\{E^{(\mathrm{a})},E^{(\mathrm{e})}\}&, 
& w\leq 1.22\,\text{fm}\,,\\
\min\{E^{(\mathrm{a})},E^{(\mathrm{e''})}\}&,
& w \geq 1.22\,\text{fm}\,\,.
\end{array} \right.
\end{equation}
Here we have used the fact that $w=1.22\,\text{fm}$ is a solution of the equation $E^{(\mathrm{e})}(w)=E^{(\mathrm{e}'')}$ for our chosen parameter set. In addition to  the examples shown in Fig.\ref{EB19054}, an example of the energies of the string configurations for $w>1.22\,\text{fm}$ is shown in Fig.\ref{EB145}.

Finally, we draw an interesting conclusion that follows from \eqref{V0w} and \eqref{V0wB}: at sufficiently large widths, the ground state potential becomes independent of the ordering. 

\subsection{Collinear geometry}

\begin{figure}[htbp]
\centering
\includegraphics[width=8.5cm]{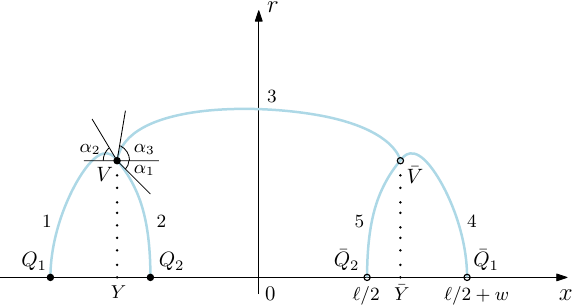}
\caption{{\small A symmetric tetraquark configuration. The axis of symmetry coincides with the $r$-axis. The distances between the quark sources are $\vert Q_1Q_2\vert=\vert\bar Q_1\bar Q_2\vert=w$ and $\vert Q_2\bar Q_2\vert=\ell$. }}
\label{cg}
\end{figure}

We now turn to the collinear geometry, with the goal of estimating the string breaking distance. So, we focus on the tetraquark configuration in which the separation between the pairs of quark sources is sufficiently large for string breaking to occur. For type-A ordering, the corresponding configuration is shown in Fig.\ref{cg}.
Here the tangent angle $\alpha_1$ is chosen to be negative, whereas the others are chosen to be positive. The reason for this is explained later.

Using the formulas of Appendix B in \cite{a4Q}, one can derive the parametric expressions for the string energies and length along the $x$-axis. When combined with the energy of the vertex, they allows us to write  

\begin{equation}\label{colgc}
w=\frac{1}{\sqrt{\s}}\Bigl({\cal L}^-(\lambda,v)+{\cal L}^+(\alpha_2,v)\Bigr)
\,,
\end{equation}
for the geometrical constraint, and 

\begin{equation}\label{colE}
\ell=\frac{1}{\sqrt{\s}}\Bigl(
{\cal L}(\bar\lambda,v)-2{\cal L}^+(\alpha_2,v)
\Bigr)
\,,\qquad
E^{(\mathrm{c})}=2\g\sqrt{\s}\Bigl(
{\cal E}^-(\lambda,v)+{\cal E}^+(\alpha_2,v)+\oh{\cal E}(\bar\lambda,v)+
3\k\frac{\ep^{-2v}}{\sqrt{v}}\Bigr)
+4c
\,.
\end{equation}
for the length $\ell$ and the energy of the configuration. Here we use the same shorthand notation as in Appendix C: $\alpha_1=\alpha$ and $\alpha_3=\bar\alpha$ for the tangent angles, along with $\lambda$ and $\bar\lambda$. In addition, relations between these angles follow from the force balance equation \cite{a4Q}. They read 

\begin{equation}\label{colfbe}
\cos\alpha-\cos\alpha_2-\cos\bar\alpha=0\,,
\qquad
\sin\alpha+\sin\alpha_2-\sin\bar\alpha=3\k(1+4v)\ep^{-3v}
\,
\end{equation}
and correspond to the $x$- and $r$-components of that equation. Using \eqref{colgc} together with \eqref{colfbe}, we can express the angles as functions of $v$, which allow us to write the energy as a function of the length $\ell$ in parametric form $\ell=\ell(v)$ and $E=E^{(\mathrm{c})}(v)$. 

For our purposes, we need to consider the large-$\ell$ limit, which corresponds to $\bar\lambda\to 1$. This is not the true IR limit but the diquark limit, in which only string $3$ becomes infinitely long, while the others remain of finite length. In this case, the tangent angle $\bar\alpha$ is expressed in terms of $v$ as $\cos\bar\alpha=v\ep^{1-v}$ \cite{a4Q}. Combining this expression with that obtained from the force balance equation and geometrical constraint, we obtain the equation $\bar\alpha(v,w)=\arccos v\ep^{1-v}$. Its solution, $\vo$, represents the upper bound on $v$. It follows from Eqs.\eqref{iLy=1} and \eqref{Ey=1} that the singular terms in $\ell$ and $E^{(\mathrm{c})}$ are 

\begin{equation}\label{colEldi}
\ell=
-\frac{1}{\sqrt{\s}}\ln(1-\bar\lambda)\,+O(1)\,,
\qquad
E^{(\mathrm{c})}=
-\g\ep\sqrt{\s}\ln(1-\bar\lambda)\,+O(1)
\,.
\end{equation}
From this, one immediately obtains

\begin{equation}\label{colEldi2}
E^{(\mathrm{c})}=\sigma\ell\,+O(1)
\,,
\end{equation}
as expected in the diquark limit.

To determine the constant term, it suffices to take the limit $\bar\lambda\to 1(v\to\vo)$ in the expression

\begin{equation}\label{colEldi3}
E^{(\mathrm{c})}-\sigma\ell=
2\g\sqrt{\s}\Bigl(
{\cal E}^-(\lambda,v)+{\cal E}^+(\alpha_2,v)+\ep{\cal L}^+(\alpha_2,v)
+\oh\bigl({\cal E}(\bar\lambda,v)-\ep{\cal L}(\bar\lambda,v)\bigr)
+3\k\frac{\ep^{-2v}}{\sqrt{v}}
\Bigr)
+4c
\,.
\end{equation}
With the help of Eq.\eqref{JLE}, we find that 

\begin{equation}\label{colEldi4}
E^{(\mathrm{c})}-\sigma\ell=
2\g\sqrt{\s}\Bigl(
{\cal E}^-(\lambda,\vo)
+{\cal E}^+(\alpha_2,\vo)+\ep{\cal L}^+(\alpha_2,\vo)
-\oh{\cal J}(\vo)
+3\k\frac{\ep^{-2\vo}}{\sqrt{\vo}}
\Bigr)
+4c
\,.
\end{equation}
Here $\cos\alpha=\oh a+\oh b\sqrt{\frac{4}{a^2+b^2}-1}\big|_{v=\vo}$ and $\cos\alpha_2=\cos\alpha-a\vert_{v=\vo}$, with $a(v)=v\ep^{1-v}$ and $b(v)=3\k(1+4v)\ep^{-3v}+\sqrt{1-a^2(v)}$. Thus,

\begin{equation}\label{colEldi5}
E^{(\mathrm{c})}
=\sigma\ell+C^{(\mathrm{c})}+o(1)\,,
\quad\text{with}\quad
C^{(\mathrm{c})}=2\g\sqrt{\s}\Bigl(
{\cal E}^-(\lambda,\vo)
+{\cal E}^+(\alpha_2,\vo)+\ep{\cal L}^+(\alpha_2,\vo)
-\oh{\cal J}(\vo)
+3\k\frac{\ep^{-2\vo}}{\sqrt{\vo}}
\Bigr)
+4c
\,. 
\end{equation}

It is noteworthy that this limit is meaningful only when the width exceeds a limiting value $\w$. As also explained in Appendix C for the rectangular geometry, this value is determined by the condition $\alpha=0$. For smaller widths, the tangent angle $\alpha$ becomes positive, and the configuration therefore changes its shape (see also \cite{astbr3Q}). One way to find $\w$ is to substitute $\alpha=0$ and $\bar\alpha=\arccos v\ep^{1-v}$ into Eqs.\eqref{colfbe} and then use the resulting solutions to evaluate the geometrical constraint. In doing so, it is useful to note that $\lambda=v$ at $\alpha=0$.  A simple numerical calculation then reproduces the expected value $\w=0.087\,\text{fm}$ found for the $QQQ$ system in \cite{astbr3Q}. Thus, for the range of $w$ chosen in Sec.III, the tangent angle $\alpha$ must be negative. 

For sufficiently large $\ell$, the string stretched between the baryon vertices breaks as the two-baryon configuration ($\mathrm{c}'$) becomes the lowest-energy configuration. Physically, this corresponds to a strong decay of the tetraquark system into a pair of doubly heavy baryons, as discussed in Sec.III. Here, we again define the string breaking distance by Eq.\eqref{l4Q}, with $\vc$ being a solution of the equation $E^{(\mathrm{c})}(v)=E^{(\mathrm{c}')}(w)$. It is instructive to make some estimates and compare them with those of Table I. The results are summarized in Table II. 
\begin{table*}[htbp]
\renewcommand{\arraystretch}{2}
\centering 	\scriptsize
\begin{tabular}{lccccccr}				
\hline
$w\,(\text{fm}) $ ~&~~ $E_{\QQq}\,(\text{GeV}) $  ~~&~~ $\ell_c\,(\text{fm})$ ~~~~&~~~~~ $\boldsymbol{\ell}_{\4Q}\,(\text{fm})$ ~&~~~~~ $\boldsymbol{\ell}_{\4Q}/\boldsymbol{\ell}_{\QQb}$
\rule[-3mm]{0mm}{8mm}
\\
\hline \hline
0.19  &    1.846  &  1.072  &~  1.113  &~~~  0.912 \\
0.31  &    1.990  &  1.014  &~  1.032  &~~~ 0.846 \\
0.37  &    2.057  &  1.000  &~  1.012  &~~~  0.830 \\
0.39  &    2.080  &  0.997  &~  1.008  &~~~  0.826 \\  
0.46  &    2.158  &  0.990  &~  0.996  &~~~  0.816 \\
0.54  &    2.246  &  0.985  &~  0.989  &~~~  0.810 \\
 \hline \hline
\end{tabular}
\caption{ \small Some estimates of the string breaking distance $\boldsymbol{\ell}_{\4Q}$ for the collinear geometry. Here $\ell_c=\ell(\vc)$ and $\boldsymbol{\ell}_{\QQb}=1.22\,\text{fm}$.}
\label{estimates2}
\end{table*}
Our main conclusion from these estimates is the same as before: the string breaking distance being dependent of the quark positions is not universal and smaller than $\boldsymbol{\ell}_{\QQb}$. It approaches the latter in the diquark limit $w\to 0$. In addition, $\boldsymbol{\ell}_{\4Q}$ is also smaller than $\boldsymbol{\ell}_{\3Q}$ obtained for the collinear geometry in \cite{astbr3Q}. 

Finally, we make a couple of remarks. First, as in Sec.III, we have considered the critical length $\ell_c=\ell(\vc)$ associated with configurations (c) and ($\mathrm{c}'$). Its definition is model-independent that facilitates future comparison with lattice QCD. Second, the equation $E^{(\mathrm{c})}(\ell)=2 E_{\QQq}$ can be solved approximately for large $\ell$. The solution is given by Eq.\eqref{lc}, but with the constant term $C^{(\mathrm{c})}$ taken from Eq.\eqref{colEldi5}. This approximation turns out to be rather accurate: it differs from the exact solution only in the fourth decimal place. 
\subsection{The potential models}

Given an interquark potential, one can in principle compute masses for a tetraquark system by solving the Schr\"odinger equation with this potential. In practice, however, this amounts to a formidable four-body problem. One common way to handle it is to take the diquark approach (approximation), which reduces the problem to two successive two-body problems. First, one assumes that quarks (antiquarks) in a color triplet (antitriplet) bind into a diquark (antidiquark).\footnote{The notion of a diquark was already proposed in the pioneer paper \cite{GM}.} Second, the tetraquark system is then treated essentially as a diquark-antidiquark bound state, using the same potential as in the quark-antiquark case. Although the estimates for the mass of $X(6400)$ (i.e., the ground state energy) appear quite plausible (see Table II in \cite{anwar}), their accuracy remains questionable if confronted with a larger body of experimental data. 

In the language of string models, the diquark approach corresponds to retaining only the tetraquark string configuration (c), alongside $(\mathrm{c}')$, and discarding all others. It is then clear that such an approximation may miss a substantial part of the physics, since the two meson configurations (a) and (e) are always relevant for the ground-state potential $V_0$. As a consequence, it may miss the decays into mesons like those in \eqref{2meson-decay} and \eqref{4meson-decay}. Moreover, the tetraquark configuration does not exist if $w$ is sufficiently large for type-A ordering, or for any $w$ in the case of type-B ordering. It is therefore perhaps not surprising that quark models yield more accurate estimates of the mass of $X(6400)$ \cite{anwar}.\footnote{Nevertheless, it remains unclear why the diquark approach is still effective in describing the masses of the known all-charm tetraquarks.} While most phenomenological models employ the diquark approach, there have also been attempts to tackle the full four-body problem directly. We mention just two works that overlap with our findings. In \cite{asadi}, the four-body problem was considered for the Schr\"odinger equation with the potential motivated by the lattice calculations \cite{suga}. The limitation of that study is its restriction to two spatial dimensions and to a fixed topology of the heavy quarks (type-A ordering). Subsequently, these issues were addressed in \cite{jido}, but using a pairwise potential. Thus, solving the Schr\"odinger equation with a genuine four-body potential remains challenging.

Finally, we should stress that all phenomenological models in the literature share a common limitation: they do not accurately take into account possible effects due to string breaking. Such effects may become important when theoretical predictions are compared with high-precision experimental data, especially for excited states. 

\section{Discussion}
\renewcommand{\theequation}{6.\arabic{equation}}
\setcounter{equation}{0}

Before solving the Schr\"odinger equation in a potential describing interquark interactions, one must have such a potential in his disposal. String models can, in principle, provide it. However, several formidable problems must be addressed. First, one must compute the off-diagonal elements of the model Hamiltonian that describe mixing among the string configurations. Second, the above analysis must be extended to QCD with $N_f=2+1$ flavors. Third, the resulting predictions must be compared with experimental data in order to single out the most realistic model. Once the Schr\"odinger equation has been solved, it may reveal the origin of the known tetraquark states and also predict new states. We are still at an early stage of this program, and much work remains to be done.

On the other hand, the physics of hadrons, in particular hadron spectroscopy, provides a natural setting in which to test and refine ideas of string theory, especially those concerning branes, in which the string junction is regarded as a brane \cite{witten}. As discussed in Sec.III, one type of string interactions admits a natural interpretation as a creation/annihilation of a brane-antibrane pair. Other examples include configuration (d), which assumes the existence of a brane-antibrane bound state, and the higher baryon vertices of Sec.V, which can be regarded as brane excitations. 

The list of known exotic hadrons continues to grow \cite{LHC}. At some point, excited tetraquarks, rather than ground state tetraquarks, are also likely to be observed. This arises the natural question of how to determine the first excited potential $V_1$. Our discussion in Sec.V suggests that its computation is another formidable task. Lattice QCD could contribute substantially to this effort, and hopefully it will be feasible in the near future. 

\begin{acknowledgments}
We would like to thank Hideo Suganuma for useful correspondence and Alex Nikolaev for help with Mathematica. This work was conducted as part of Program FFWR-2024-0011 at the Landau Institute.
\end{acknowledgments}

 \appendix
 \section{Notation and useful formulas}
\renewcommand{\theequation}{A.\arabic{equation}}
\setcounter{equation}{0}

Throughout this paper, we denote heavy quarks (antiquarks) by $Q(\bar Q)$, light quarks (antiquarks) by $q(\bar q)$, and baryon (antibaryon) vertices by $V(\bar V)$. Unless otherwise specified, heavy quark sources are located on the boundary of the five-dimensional space \eqref{metric}, at $r=0$, whereas light quarks and vertices are placed in the interior at $r=\rq$ and $r=\rv$, respectively. For convenience, we introduce the dimensionless variables $q=\s\rq^2$ and $v=\s\rv^2$. These variables measure the proximity of the objects to the soft-wall, located at $1$ in these units. 

To write the resulting formulas in a compact form, we use the basic functions introduced in \cite{astbr3Q}, together with several newly defined ones:

(i) The function ${\cal L}^+$ is given by  

\begin{equation}\label{fL+}
{\cal L}^+(\alpha,x)=\cos\alpha\sqrt{x}\int^1_0 du\, u^2\, \ep^{x (1-u^2)}
\Bigl[1-\cos^2{}\hspace{-1mm}\alpha\, u^4\ep^{2x(1-u^2)}\Bigr]^{-\frac{1}{2}}
\,,
\qquad
0\leq\alpha\leq\frac{\pi}{2}\,,
\qquad 
0\leq x\leq 1
\,.
\end{equation}
This non-negative function vanishes at $\alpha=\frac{\pi}{2}$ and at $x=0$. For small $x$, if $\alpha$ approaches $\alpha_0$, it behaves as 

\begin{equation}\label{fL+smallx}
{\cal L}^+(\alpha,x)=\sqrt{x}\bigl({\cal L}^+_0(\alpha_0)+O(x)\bigr)
\,,
\qquad\text{with}\qquad
{\cal L}^+_0(\alpha_0)=\frac{1}{4}
\cos^{-\oh}\hspace{-.9mm}\alpha_0\,B\bigl(\cos^2\hspace{-.9mm}\alpha_0;\tfrac{3}{4},\tfrac{1}{2}\bigr)
\,,
\end{equation}
where $B(z;a,b)$ denotes the incomplete beta function. It has a logarithmic singularity at $(0,1)$. So, 

\begin{equation}\label{fL+x1}
{\cal L}^+(\alpha,x)=-\oh\ln(1-x)+O(1)
\,,
\end{equation}
as $\alpha\rightarrow 0$ and $x\to 1$. For brevity, we write ${\cal L}^+(x)$ instead of ${\cal L}^+(0,x)$.

(ii) The function ${\cal L}^-$ is given by 

\begin{equation}\label{fL-}
{\cal L}^-(y,x)=\sqrt{y}
\biggl(\,
\int^1_0 du\, u^2\, \ep^{y(1-u^2)}
\Bigl[1-u^4\,\ep^{2y(1-u^2)}\Bigr]^{-\frac{1}{2}}
+
\int^1_
{\sqrt{\frac{x}{y}}} 
du\, u^2\, \ep^{y(1-u^2)}
\Bigl[1-u^4\,\ep^{2y(1-u^2)}\Bigr]^{-\frac{1}{2}}
\,\biggr)
\,,
\quad
0\leq x\leq y\leq 1
\,.
\end{equation}
It is non-negative and vanishes at the origin. For $y=x/\rho$, with non-zero $\rho$ as $x\rightarrow 0$, the small-$x$ behavior of ${\cal L}^-$ is

\begin{equation}\label{L-y=0}
	{\cal L}^-(y,x)=\sqrt{x}\bigl({\cal L}^-_0(\rho)+O(x)\bigr)
	\,,\qquad\text{with}\qquad
	{\cal L}^-_0(\rho)=\frac{1}{4}\rho^{-\oh}{\cal B}\bigl(1-\rho^2;\tfrac{1}{2},\tfrac{3}{4}\bigr)
\,.
\end{equation}
Here ${\cal B}(z;a,b)=B(a,b)+B(z;a,b)$. This function also has a logarithmic singularity at $y=1$, where

\begin{equation}\label{L-y=1}
{\cal L}^-(y,x)=-\ln(1-y)+O(1)\,,
\quad
\text{with fixed}\,\,\,x
\,.
\end{equation}
The functions ${\cal L}^{\pm}$ are related by ${\cal L}^+(x)={\cal L}^-(x,x)$. 

(iii) The function ${\cal L}$ is defined as

\begin{equation}\label{iL}
{\cal L}(y,x)=2\sqrt{y}
\int^1_
{\sqrt{\frac{x}{y}}} 
du\, u^2\, \ep^{y(1-u^2)}
\Bigl[1-u^4\,\ep^{2y(1-u^2)}\Bigr]^{-\frac{1}{2}}
\,,
\qquad 
0<x\leq y
\,.
\end{equation}
It is a non-negative function that vanishes at $y=0$ and $x=y$. For $y=x/\rho$, with non-zero $\rho$ as $x\rightarrow 0$, the small-$x$ behavior is

\begin{equation}\label{iLy=0}
{\cal L}(y,x)
=
\sqrt{x}
\bigl({\cal L}_0(\rho)+O(1)\bigr)\,,
\qquad\text{with}\qquad
	{\cal L}_0(\rho)=\frac{1}{2}\rho^{-\oh}B\bigl(1-\rho^2;\tfrac{1}{2},\tfrac{3}{4}\bigr)
\,.	
\end{equation}
The function ${\cal L}$ is singular at $y=1$, where

\begin{equation}\label{iLy=1}
{\cal L}(y,x)=-\ln(1-y)+O(1)\,,
\quad
\text{with}\,\,\,\text{fixed}\,\,\, x
\,.
\end{equation}
The ${\cal L}$'s are related by the identity

\begin{equation}\label{LL-}
	{\cal L}(y,x)={\cal L}^-(y,x)-{\cal L}^+(\alpha,x)
	\,,
\end{equation}
where $\cos\alpha=\frac{x}{y}\ep^{y-x}$.

(iv) The function ${\cal E}^+$ is given by  

\begin{equation}\label{fE+}
{\cal E}^+(\alpha,x)=
\frac{1}{\sqrt{x}}
\int^1_0\,\frac{du}{u^2}\,\biggl(\ep^{x u^2}
\Bigl[
1-\cos^2{}\hspace{-1mm}\alpha\,u^4\ep^{2x (1-u^2)}
\Bigr]^{-\frac{1}{2}}-1-u^2\biggr)
\,,
\qquad
0\leq\alpha\leq\frac{\pi}{2}\,,
\qquad 
0\leq x\leq 1
\,.
\end{equation}
It is singular at $x=0$ and at $(0,1)$. If $\alpha$ approaches $\alpha_0$ as $x\rightarrow 0$, its small-$x$ behavior is 

\begin{equation}\label{fE+smallx}
{\cal E}^+(\alpha,x)=\frac{1}{\sqrt{x}}\bigl({\cal E}^+_0(\alpha_0)+O(x)\bigr)
\,,
\qquad\text{with}\qquad
{\cal E}_0^+(\alpha_0)=
\frac{1}{4}
\cos^{\oh}\hspace{-.9mm}\alpha_0\,B\bigl(\cos^2\hspace{-.9mm}\alpha_0;-\tfrac{1}{4},\tfrac{1}{2}\bigr)
\,.
\end{equation}
For $\alpha\rightarrow 0$ and $x\to 1$, the function behaves as

\begin{equation}\label{E-x=1}
	{\cal E}^+(\alpha,x)=-\oh\ep\ln(1-x)+O(1)
	\,.
\end{equation}
We simplify notation by writing  ${\cal E}^+(x)$ instead of ${\cal E}^+(0,x)$.

(v) The ${\cal E}^-$ function is given by 

\begin{equation}\label{fE-}
{\cal E}^-(y,x)=\frac{1}{\sqrt{y}}
\biggl(
\int^1_0\,\frac{du}{u^2}\,
\Bigl(\ep^{y u^2}\Bigl[1-u^4\,\ep^{2y(1-u^2)}\Bigr]^{-\frac{1}{2}}
-1-u^2\Bigr)
+
\int^1_{\sqrt{\frac{x}{y}}}\,\frac{du}{u^2}\,\ep^{y u^2}
\Bigl[1-u^4\,\ep^{2y(1-u^2)}\Bigr]^{-\frac{1}{2}}
\biggr) 
\,,
\,\,\,
0\leq x\leq y\leq 1
\,.
\end{equation}
${\cal E}^-$ is singular at $(0,0)$ and at $y=1$. More specifically, for $y=x/\rho$ with non-zero $\rho$, as $x\rightarrow 0$, its small-$x$ behavior is

\begin{equation}\label{E-y=0}
	{\cal E}^-(y,x)=\frac{1}{\sqrt{x}}\Bigl({\cal E}^-_0(\rho)+O(x)\Bigr)
	\,,\qquad\text{with}\qquad
	{\cal E}^-_0(\rho)=\frac{1}{4}\rho^{\oh}{\cal B}\bigl(1-\rho^2;\tfrac{1}{2},-\tfrac{1}{4}\bigr)
\,.
\end{equation}
On the other hand, near $y=1$ with fixed $x$, it behaves as

\begin{equation}\label{E-y=1}
	{\cal E}^-(y,x)=-\ep\ln(1-y)+O(1)
	\,.
\end{equation}
The functions ${\cal E}^+$ and ${\cal E}^-$ are related by ${\cal E}^+(x)={\cal E}^-(x,x)$.

(vi) The ${\cal E}$ function is defined by 

\begin{equation}\label{ifE-}
{\cal E}(y,x)=
\frac{2}{\sqrt{y}}
\int^1_{\sqrt{\frac{x}{y}}}\,\frac{du}{u^2}\,
\ep^{y u^2}
\Bigl[1-u^4\ep^{2y(1-u^2)}
\Bigr]^{-\frac{1}{2}}
\,,
\qquad
0<x\leq y
\,.
\end{equation}
It is singular at the origin. For $y=x/\rho$, with non-zero $\rho$ as $x\rightarrow 0$, the small-$x$ behavior of ${\cal E}$ is

\begin{equation}\label{iEx=0}
	{\cal E}(y,x)=\frac{1}{\sqrt{x}}\Bigl({\cal E}_0(\rho)+o(x)\Bigr)
	\,,\qquad\text{with}\qquad
	{\cal E}_0(\rho)=\frac{1}{2}\rho^{\oh}B\bigl(1-\rho^2;\tfrac{1}{2},-\tfrac{1}{4}\bigr)\,.
\end{equation}
It is also singular at $y=1$, where it has a logarithmic singularity, 

\begin{equation}\label{Ey=1}
	{\cal E}(y,x)=-\ep\ln(1-y)+O(1)
	\,,\quad\text{with}\,\,\,\text{fixed}\,\,\,x\,.
\end{equation}
The ${\cal E}$'s satisfy the identity 

\begin{equation}\label{EE-}
	{\cal E}(y,x)
	={\cal E}^-(y,x)
	-
	{\cal E}^+(\alpha,x)
	\,,
\end{equation}
with $\cos\alpha=\frac{x}{y}\ep^{y-x}$.

(vii) The function ${\cal Q}$ is given by 

\begin{equation}\label{Q}
{\cal Q}(x)=\sqrt{\pi}\text{erfi}(\sqrt{x})-\frac{\ep^x}{\sqrt{x}}
\,,
\qquad
0<x\leq 1
\,,
\end{equation}
where $\text{erfi}(x)$ denotes the imaginary error function. This is a special case of ${\cal E}^+$ at $\alpha=\frac{\pi}{2}$. Its small-$x$ behavior is  

\begin{equation}\label{Q0}
{\cal Q}(x)=-\frac{1}{\sqrt{x}}+\sqrt{x}+O(x^{\frac{3}{2}})
\,.
\end{equation}

(viii) The function ${\cal I}$ is 

\begin{equation}\label{I}
	{\cal I}(x)=
	I_0
	-
	\int_{\sqrt{x}}^1\frac{du}{u^2}\ep^{u^2}\Bigl[1-u^4\ep^{2(1-u^2)}\Bigr]^{\frac{1}{2}}
	\,,
\quad\text{with}\quad 
I_0=\int_0^1\frac{du}{u^2}\Bigl(1+u^2-\ep^{u^2}\Bigl[1-u^4\ep^{2(1-u^2)}\Bigr]^{\frac{1}{2}}\Bigr)
\,,
\quad
0< x\leq 1
\,.
\end{equation}
A special case is ${\cal I}(1)=I_0$. Numerically, $I_0=0.751$. This function is related to the ${\cal L}^-$ and $ {\cal E}^-$ functions by 

\begin{equation}\label{ILE}
	{\cal I}(x)=\ep\,{\cal L}^-(y,x)-{\cal E}^-(y,x)
	\quad\text{as}\,\,\, y\rightarrow 1\,\,\text{at}
	\,\,\,\text{fixed}\,\,\,x\,.
	\end{equation}

(ix) The ${\cal J}$ function is defined by 

\begin{equation}\label{iI}
	{\cal J}(x)
	=2\bigl({\cal I}(x)-I_0\bigr)
	=
	-2
	\int_{\sqrt{x}}^1\frac{du}{u^2}\ep^{u^2}\Bigl[1-u^4\ep^{2(1-u^2)}\Bigr]^{\frac{1}{2}}
	\,,
\quad
0< x\leq 1
\end{equation}
Like ${\cal I}$, it  satisfies the relation 

\begin{equation}\label{JLE}
	{\cal J}(x)=\ep\,{\cal L}(y,x)-{\cal E}(y,x)
	\quad\text{as}\,\,\, y\rightarrow 1\,\,\text{at}
	\,\,\,\text{fixed}\,\,\,x\,.
	\end{equation}
\section{Some details on the $Q\bar Q$ and $QQq$ quark systems}
\renewcommand{\theequation}{B.\arabic{equation}}
\setcounter{equation}{0}

In this Appendix we give a brief summary of the basic results about the B-O potentials of the $Q\bar Q$ and $QQq$ systems. These results are pertinent to our discussion in Secs. III and V. For standard explanations, see \cite{az1,a-strb,ahybrid,aQQq} whose conventions we follow, unless otherwise stated.
\subsection{The $Q\bar Q$ system}

\begin{figure}[H]
\centering
\includegraphics[width=12.75cm]{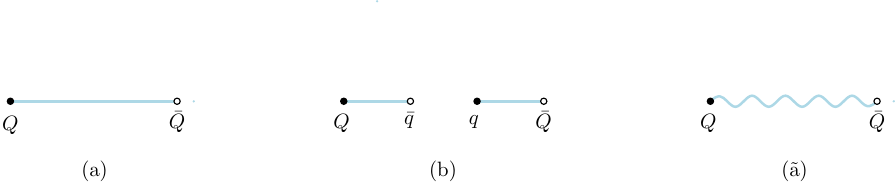}
\caption{{\small Some string configurations for the $Q\bar Q$ system in four dimensions.}}
\label{4QQb}
\end{figure}

We begin with the quark-antiquark pair in the presence of two light flavors of equal mass. For the purposes of the present paper we need only a few string configurations, which are shown in Fig.\ref{4QQb}. Configuration (a) is the simplest
\noindent one, consisting of a valence quark and an antiquark connected by a string. Configuration (b) is disconnected. It is obtained by adding a pair of light quarks and attaching strings to them in such a way that two heavy-light mesons are formed. These configurations have a physical meaning: in string models, the decay of a heavy meson, $Q\bar Q\rightarrow Q\bar q+q\bar Q$, is described  as a transition between them. The transition occurs through string breaking. Configuration $(\tilde{\mathrm{a}})$ is an extension of configuration (a) in which the string is in its excited state.  

\begin{figure}[htbp]
\centering
\includegraphics[width=17.25cm]{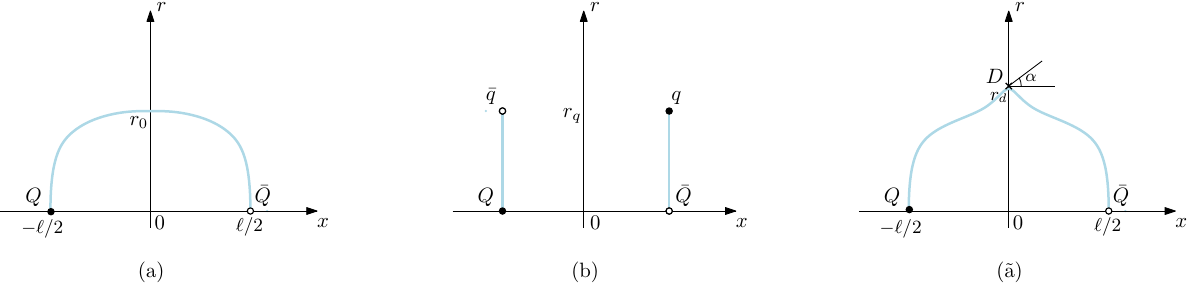}
\caption{{\small The five-dimensional counterparts to the string configurations of Fig.\ref{4QQb}. The heavy quarks are separated by a distance $\ell$ and are located at $r=0$, while the light quarks and defect lie in the bulk at $r=\rq$ and at $r=r_d$, respectively. In (a), $r_0$ is the $r$-coordinate of the string turning point. In $(\tilde{\mathrm{a}})$, we specialize to the string excitation associated with the hybrid potential $\Sigma_u^-$ \cite{ahybrid}. A point-like defect $D$ leads to the cusp formation at $x=0$ with angle $\alpha$.}}
\label{5QQb}
\end{figure}

In five dimensions, the configurations are constructed from strings attached to the heavy quark sources on the boundary of the five-dimensional space \eqref{metric}, as shown in Fig.\ref{5QQb}. In addition, configurations (b) and ($\tilde{\mathrm{a}}$) contain 
point-like objects in the bulk: a pair of light quarks and a defect, respectively. 

Configuration (a) was originally discussed in the context of the AdS/CFT correspondence \cite{malda}. Apart from the renormalization scheme, essentially the same technique applies to the geometry \eqref{metric} \cite{az1}. In this case, the relation between the quark separation distance and the string energy is given in parametric form by\footnote{Throughout this Appendix, we omit the superscripts associated with the configurations to reduce clutter.}

\begin{equation}\label{EQQb}
\ell= \frac{2}{\sqrt{\s}}{\cal L}^+(\lambda)
 \,,
\quad
E_{\QQb}=2\g\sqrt{\s}\,{\cal E}^+(\lambda)+2c\,,
\end{equation}
where $\lambda=\s \ro^2$ is a dimensionless parameter running from $0$ to $1$, and $c$ is a normalization constant. The functions ${\cal L}^+$ and ${\cal E}^+$ are defined  in Appendix A. 

For future reference, we note that the small-$\ell$ behavior of $E_{\QQb}$ is given by 

\begin{equation}\label{EQQb-small}
E_{\QQb}(\ell)=-\frac{\alpha_{\QQb}}{\ell}+2c+\boldsymbol{\sigma}_{\QQb}\ell +o(\ell)
\,,
\end{equation}
with 
\begin{equation}\label{alpha-QQb}
\alpha_{\QQb}=(2\pi)^3\Gamma^{-4}\bigl(\tfrac{1}{4}\bigr)\g
	\,,\qquad
	\boldsymbol{\sigma}_{\QQb}=\frac{1}{8\pi^2}\Gamma^{4}\bigl(\tfrac{1}{4}\bigr)\g\s
	\,.
\end{equation}

On the other hand, the large-$\ell$ behavior is linear:  

\begin{equation}\label{EQQb-large}
E_{\QQb}(\ell)=\sigma\ell+C_{\QQb}+o(1)
\,,\qquad
\text{with}
\qquad
\sigma=\ep\g\s \qquad\text{and}\qquad C_{\QQb}=2c-2\g\sqrt{\s}\,I_0
\,.
\end{equation}
Here $I_0$ is defined in Appendix A, and $\sigma$ is the string tension. It has a larger value than the coefficient $\boldsymbol{\sigma}_{\QQb}$ in \eqref{EQQb-small}. Numerically, the ratio of $\boldsymbol{\sigma}_{\QQb}$ to $\sigma$ is approximately $0.805$.

The disconnected configuration (b) can be interpreted as a hadronic molecule composed of two heavy-light mesons. Assuming that the mesons are nearly non-interacting, the total energy is simply twice the energy $E_{\Qqb}$. The latter was computed in \cite{a-strb}, with the result

\begin{equation}\label{Qqb}
E_{\Qqb}=\g\sqrt{\s}\Bigl({\cal Q}(q)+\n \frac{\ep^{\oh q}}{\sqrt{q}}\Bigr)+c
\,,
	\end{equation}
where the function ${\cal Q}$ is given in Appendix A, $c$ is the normalization constant as before, and $q=\s\rq^2$ is a solution of the equation 

\begin{equation}\label{q}
\n(q-1)+\ep^{\frac{q}{2}}=0
\,
\end{equation}
in the interval $[0,1]$. Numerically, $q=0.566$. This equation is nothing else but the $r$-component of the force balance equation. It can be derived by varying the action $S=S_{\text{\tiny NG}}+S_{\text q}$ with respect to $\rq$. 

So far we have considered unexcited strings. However, a QCD string also has excited states, which are related to the hybrid heavy quark potentials. These potentials have been studied in lattice QCD. By contrast, their string constructions remain largely unknown. In the present model, there is one for the potential $\Sigma_u^-$ \cite{ahybrid}. So, we will consider only this type of string excitations. It has a geometrical interpretation as a cusp formation on a string (see Fig.\ref{5QQb}$\tilde{\mathrm{a}})$. 

For configuration ($\tilde{\mathrm{a}}$), the relation between the quark separation distance and its energy is written parametrically as

\begin{equation}\label{EQQb-hybrid}
\ell= \frac{2}{\sqrt{\s}}{\cal L}^+(\alpha,v)
 \,,
\quad
\tilde{E}_{\QQb}=2\g\sqrt{\s}
\Bigl({\cal E}^+(\alpha,v)+\tilde{\k}\,\frac{\ep^{-2v}}{{\sqrt{v}}}\Bigr)
+
2c
\,,
\end{equation}
with the cusp angle
\begin{equation}\label{alpha}
	\sin\alpha=\tilde{\k}(1+4v)\ep^{-3v}
	\,.
\end{equation}
Here $\tilde{\k}$ is a parameter that characterizes the defect. The dimensionless parameter $v=\s r_d^2$ runs from  $v_{\min}$ to $v_{\max}$, where $v_{\min}$ and $v_{\max}$ are  solutions to the equations $\cos\alpha=0$ and $\cos\alpha=v\ep^{1-v}$, respectively.   

There is one important remark. For example, at $\tilde{\k}=2000$ \cite{ahybrid}, a simple calculation gives $v_{\min}=3.430$ and $v_{\max}=3.447$. Thus, the defect is allowed to go beyond the soft wall. Although this may seem counterintuitive, an important point is that this is an excited string. For $\alpha$ given by \eqref{alpha}, both functions ${\cal L}^+$ and ${\cal E}^+$ are well defined and monotonically increasing on the interval $[v_{\min},v_{\max}]$. They develop logarithmic singularities at $v=v_{\max}$. 

The function $\tilde{E}_{\QQb}(\ell)$ behaves as 

\begin{equation}\label{EQQb-hybrid-small}
\tilde{E}_{\QQb}(\ell)=\tilde{E}_{\QQb}(v_{\min})
+O(\ell^2)
\,,\qquad
\text{with}
\qquad
\tilde{E}_{\QQb}(v_{\min})=2\g\sqrt{\s}
\Bigl({\cal Q}(v_{\min})+\tilde{\k}\,\frac{\ep^{-2v_{\min}}}{{\sqrt{v_{\min}}}}\Bigr)
+
2c
\,,
\end{equation}
for small $\ell$ and as 
\begin{equation}\label{EQQb-hybrid-large}
\tilde{E}_{\QQb}(\ell)=\sigma\ell+\tilde{C}_{\QQb}+o(1)
\,,\qquad
\text{with}
\qquad
\tilde{C}_{\QQb}=C_{\QQb}+2\g\sqrt{\s}
\biggl(
\int_1^{\sqrt{v_{\max}}}
dv\sqrt{v^{-4}\ep^{2v^2}-\ep^2}\,+\tilde{\k}\,\frac{\ep^{-2v_{\max}}}{\sqrt{v_{\max}}}
\biggr)
\,,
\end{equation}
for large $\ell$. The asymptotic expansions \eqref{EQQb-large} and \eqref{EQQb-hybrid-large} have the same leading linear terms, whereas the constant terms are different due to the contribution of the defect. 

\begin{figure}[H]
\centering
\includegraphics[width=7.85cm]{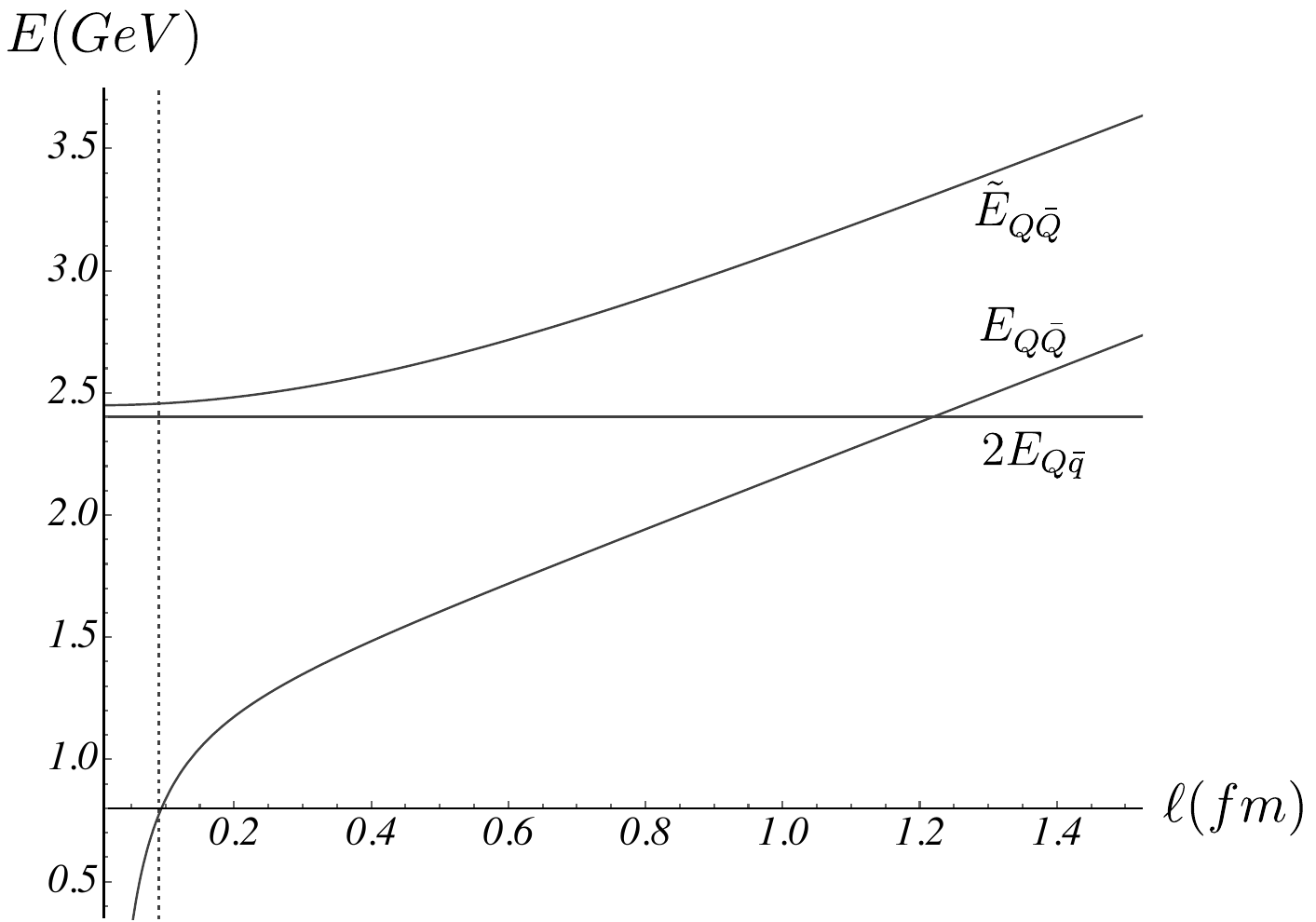}
\hspace{1cm}
\includegraphics[width=7.85cm]{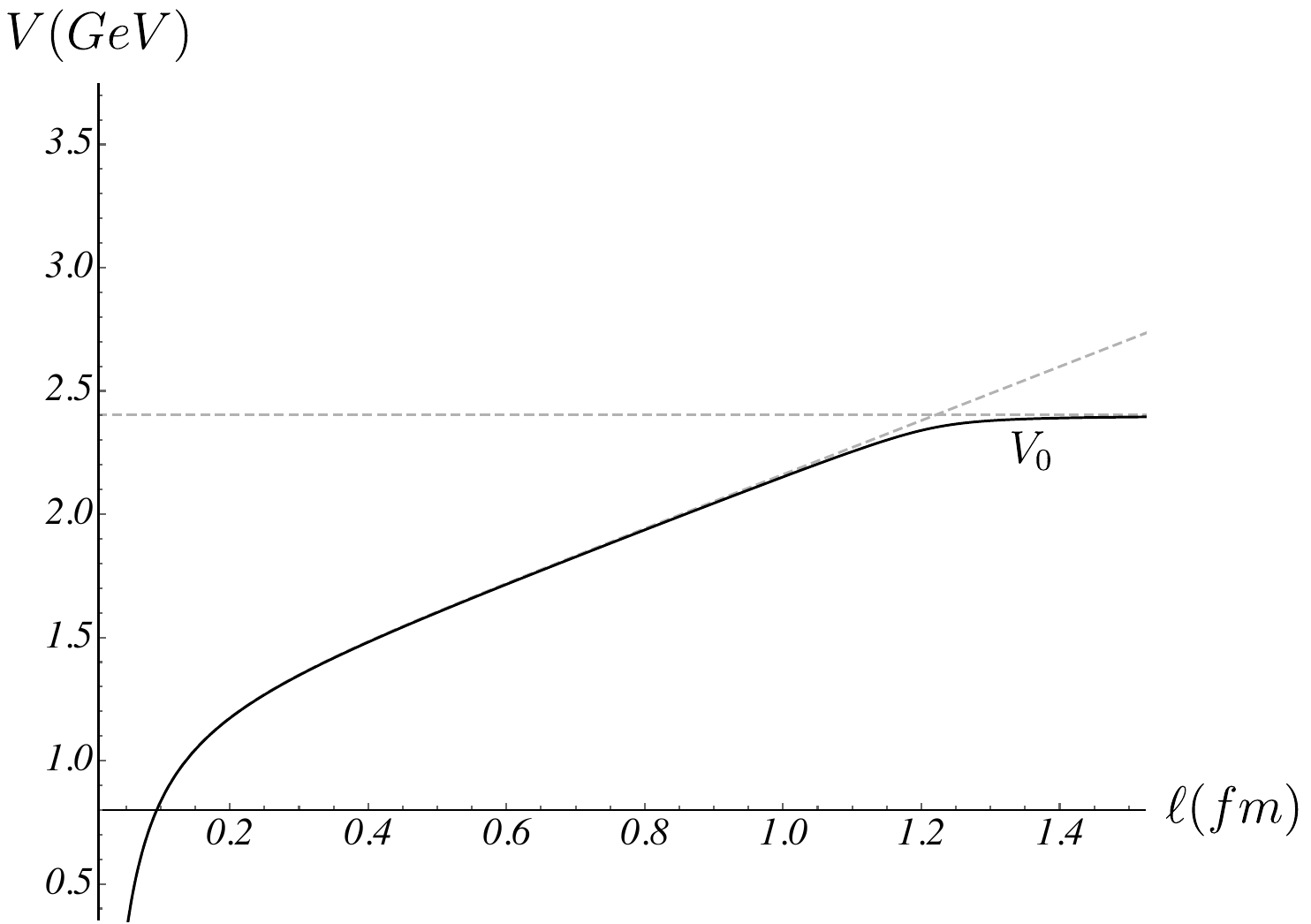}
\caption{{\small Left: The energies of the string configurations. Here we set $\tilde\k=2000$, as in \cite{ahybrid}. The vertical line is at $\ell=0.09\,\text{fm}$, corresponding to the lattice spacing in \cite{wagner}. Right: The ground state potential determined using the expression \eqref{V0QQb} with $\Theta=47\,\text{MeV}$ from \cite{bulava}. The relevant energies are shown in dashed.}}
\label{VQQb}
\end{figure}

In the left panel of Fig.\ref{VQQb}, we plot the energies of the string configurations as functions of quark separation. It is noteworthy that there is a sizable gap between $E_{\QQb}$ and $\tilde{E}_{\QQb}$ at large $\ell$. The construction of $\tilde{E}_{\QQb}$, however, requires a caveat. For small $\ell$, $\tilde{E}_{\QQb}$ approaches a constant, whereas the potential $\Sigma_u^-$ is expected to exhibit Coulomb-like behavior (see, for example, \cite{wagner}). Although the agreement with the lattice data of \cite{kuti} is quite good, the discrepancy occurs at separations below the lattice spacing, as shown in the figure. This is expected in the string model, which is not reliable at very small quark separations, where perturbative QCD becomes applicable. 

The ground state B-O potential is formally defined by $V_{0}=\min\bigl(E_{\QQb},2E_{\Qqb}\bigr)$. However, this formal definition does not accurately describe the behavior at intermediate separations. Addressing this issue through the mixing analysis based on the Hamiltonian \eqref{H} leads to technical complications. In practice, it is much simpler to truncate the infinite number of string configurations to two ones and consider a model Hamiltonian of a two-state system,

\begin{equation}\label{HD-QQb}
{\cal H}(\ell)=
\begin{pmatrix}
E_{\QQb}(\ell) & \Theta \\
\Theta & 2E_{\Qqb} \\
\end{pmatrix}
\,,
\end{equation}
where $\Theta$ describes the mixing between the states. The potential is then obtained as the smallest eigenvalue of the model Hamiltonian. Explicitly, 

\begin{equation}\label{V0QQb}
V_{0}=\oh\bigl(E_{\QQb}+2E_{\Qqb}\bigr)
-
\sqrt{\frac{1}{4}\bigl(E_{\QQb}-2E_{\Qqb}\bigr)^2+\Theta^2}
\,.	
\end{equation}

The string breaking distance is defined by equating the energies of the configurations \cite{bulava}

\begin{equation}\label{LcQQb}
E_{\QQb}(\ell_{\QQb})=2E_{\Qqb}
\,.
\end{equation}
This distance determines which configuration dominates the ground state at a given heavy quark separation. At large separations, $E_{\QQb}(\ell)$ becomes a linear function of $\ell$ and the equation simplifies considerably.\footnote{For the parameter values used here, this holds for $\ell \gtrsim 0.5\,\text{fm}$, whereas the string breaking distance is about $1\,\text{fm}$.} If so, then from Eqs.\eqref{EQQb-large} and \eqref{Qqb} it follows that the string breaking distance is  

\begin{equation}\label{LQQb-large}
\ell_{\QQb}\approx
\frac{2}{\ep\sqrt{\s}}
\Bigl(
{\cal Q}(q)+\n\frac{\ep^{\oh q}}{\sqrt{q}}+I_0
\Bigr)
\,.
\end{equation}

Finally, we illustrate this construction by plotting $V_0$ in the right panel of Fig.\ref{VQQb}. As expected, $V_0$ approaches $E_{\QQb}$ as $\ell\to 0$ and approaches $2E_{\Qqb}$ as $\ell\to\infty$. The transition between these two asymptotic regimes occurs around $\ell=1.22\,\text{fm}$, consistent with our expectations \cite{bulava}.

\subsection{The $QQq$ system}

\begin{figure}[htbp]
\centering
\includegraphics[width=2.5cm]{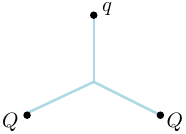}
\caption{{\small A typical string configuration for a doubly heavy baryon in four dimensions.}}
\label{4QQq}
\end{figure}

\begin{figure}[htbp]
\centering
\includegraphics[width=5.6cm]{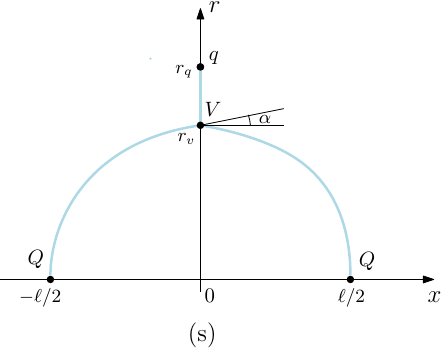}
 \hspace{2cm}
 \includegraphics[width=5.6cm]{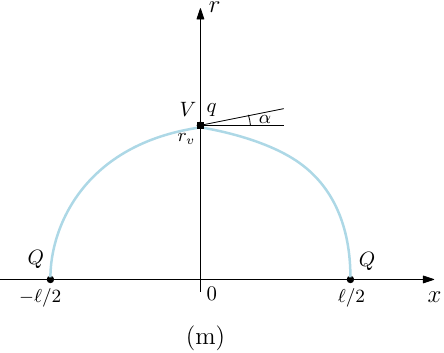}
\caption{{\small String configurations contributing to the ground state energy of the $QQq$ system at small and intermediate separations between the heavy quarks. They are labeled so that each letter denotes a particular range of heavy quark separations (for example, s stands for small). Here $\alpha$ is a tangent angle of the left string at point $V$. In configuration (s), the light quark and vertex are located at $r=r_q$ and $r=r_v$, respectively, whereas in configuration (m) both are at $r=r_v$.}}
\label{QQq}
\end{figure}

We now review some aspects of the $QQq$ quark system which are important for our discussion in Secs.III and V. In particular, we need its ground state energy as a function of the separation between the heavy quarks at small and intermediate separations. For this purpose, it suffices to consider the simple string configuration shown in Fig.\ref{4QQq}.
Curiously, its five-dimensional counterpart is more involved and consists of three configurations, only two of which are relevant here. They are shown in Fig.\ref{QQq}. 
The reason is that each configuration describes a particular range of heavy quark separations. The full range is obtained by combining all of them \cite{aQQq}.

For configuration (s), the relation between the energy and the heavy quark separation is written in parametric form as

\begin{equation}\label{EQQqs}
\ell^{\,\text{(s)}}=\frac{2}{\sqrt{\s}}{\cal L}^+(\alpha,v)
\,,
\qquad
E_{\QQq}^{\,\text{(s)}}=\g\sqrt{\s}
\Bigl(
2{\cal E}^+(\alpha,v)
+
\n\frac{\ep^{\oh q}}{\sqrt{q}}+3\k\frac{\ep^{-2v}}{\sqrt{v}}
+
{\cal Q}(q)-{\cal Q}(v)
\Bigr)
+2c\,,
\end{equation}
where the parameter $v=\s r_v^2$ varies from $0$ to $q$. The value of $q$ is determined by Eq.\eqref{q}. The parameter $\k$ characterizes the baryon vertex (see Sec.II). The tangent angle $\alpha$ is expressed in terms of $v$ through the force balance equation at the vertex, which gives

\begin{equation}\label{alpha1}
\sin\alpha=\oh\bigl(1+3\k(1+4v)\ep^{-3v}\bigr)
\,.	
\end{equation}
At this point, it is worth noting that in the limit $\ell\rightarrow 0$ or, equivalently $v\to 0$, the energy reduces to the sum $E_{\QQq}^{\,\text{(s)}}=E_{\qQb}+E_{\QQ}$, as expected from heavy quark-diquark symmetry \cite{wise}. The quark-quark potential is given explicitly by \eqref{diquarkEc2}.

At intermediate separations, the system is described by configuration (m). In this case, one string shrinks into a point, so that the position of the light quark coincides with that of the vertex. The separation distance is expressed in terms of $v$ and $\alpha$ by the same formula as before, but for another parameter range, while the energy is given by

\begin{equation}\label{EQQqm} 
E_{\QQq}^{\,\text{(m)}}=\g\sqrt{\s}
\Bigl(
2{\cal E}^+(\alpha,v)
+
\frac{1}{{\sqrt{v}}}\bigl(
\n\ep^{\oh v}+
3\k\ep^{-2v}
\bigr)
\Bigr)
+2c\,.
\end{equation}
The parameter $v$ now varies from $q$ to $\vm$, where $\vm$ is a solution to the equation 

\begin{equation}\label{v0}
\n (1-v)+3\k(1+4v)\ep^{-\frac{5}{2}v}=0
\,.
\end{equation}
The force balance equation in this case leads to  

\begin{equation}\label{alpha2}
\sin\alpha=\oh\bigl(
\n (1-v)\ep^{-\oh v}+3\k(1+4v)\ep^{-3v}
\bigr)
\,.
\end{equation}
Note that $\alpha(\vm)=0$. It is also worth noting that, for our set of the parameters, configuration (m) describes heavy quark separations up to $1.11\,\text{fm}$ that is more than sufficient for our purposes. 

\section{The basic tetraquark configurations}
\renewcommand{\theequation}{C.\arabic{equation}}
\setcounter{equation}{0}

\begin{figure}[H]
\centering
\includegraphics[width=7.25cm]{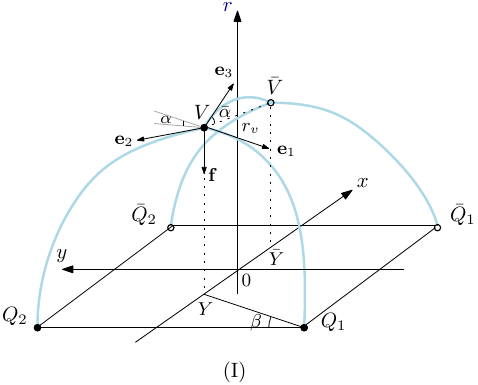}
\hspace{1.5cm}
\includegraphics[width=7.257cm]{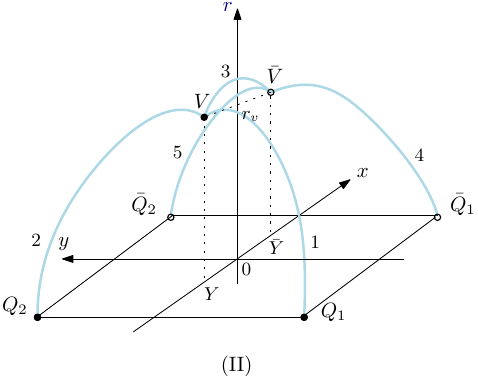}
\caption{{\small The basic tetraquark configurations in five dimensions. The heavy quark sources lie on the boundary of the five-dimensional space, at the vertices of a rectangle whose center is at the origin. The configurations are symmetric under reflections in the $x$- and $y$-axes. The baryon vertices are located at $r=\rv$ in the bulk. $Y$ and $\bar Y$ are their projections onto the boundary, and $\beta=\angle YQ_1Q_2$. The strings are labeled by $1,\dots,5$, as shown in the right panel. For the tangent angles at the vertices, we use the symmetry-based abbreviation: $\alpha_{1,2,4,5}=\alpha$ and $\alpha_3=\bar\alpha$. Left: $\alpha\geq 0$. The arrows indicate the forces acting on $V$, with the gravitational force $\mathbf{f}$ directed downward. Right: $\alpha\leq 0$.}}
\label{tetra5d}
\end{figure}

In this Appendix, we will discuss the basic tetraquark configurations of \cite{a4Q}, focusing on the new features arising from the geometrical constraint \eqref{gc}. These configurations are shown in Fig.\ref{tetra5d}. For convenience, we list the following  
\noindent simple but useful formulas:

\begin{equation}\label{lw}
\ell=\vert Q_1\bar Q_1\vert
=2\sin\beta\vert Q_1Y\vert+\vert Y\bar Y\vert
\,,\qquad
w=\vert Q_1Q_2 \vert =
2\cos\beta\vert Q_1Y\vert 
\,,
\end{equation}
which hold for $0\leq\beta\leq\arctan\frac{\ell}{w}$. 

The total action governing each tetraquark configuration is the sum of the Nambu-Goto actions for the strings and the actions for the two vertices

\begin{equation}\label{action5d}
S=\sum_{i=1}^5 S_{\text{\tiny NG}}^{(i)}\,+2S_{\text{\tiny V}}
\,.	
\end{equation}
Varying the total action with respect to the fields entering the Nambu-Goto actions yields the equations of motion for the strings, which determine their profiles (for details, see Appendix B of \cite{a4Q}). In addition, varying it with respect to the position of the vertex $V$ gives the force balance equation, whose component form is \cite{a4Q}

\begin{equation}\label{fbec}
	2\sin\beta\cos\alpha-\cos\bar\alpha=0\,,\qquad
	2\sin\alpha-\sin\bar\alpha-3\k(1+4v)\ep^{-3v}=0
	\,,
\end{equation}
with $v=\s\rv^2$. By symmetry, the same equations apply to the vertex $\bar V$, with identical tangent angles. 
\subsection{Configuration (I)}

As explained in Appendix B of \cite{a4Q}, the string lengths along the boundary are given by $\vert Q_1Y\vert=\frac{1}{\sqrt{\s}}{\cal L}^+(\alpha,v)$ and $\vert Y\bar Y\vert=\frac{1}{\sqrt{\s}}{\cal L}(\bar \lambda,v)$, where $\bar\lambda=-\text{ProductLog}(-v\ep^{-v}/\cos\bar\alpha)$ and the functions $\mathcal{L}^+$ and $\mathcal{L}$ are defined in Appendix A. Combining these expressions with those for the string energies \cite{a4Q} allows us to write the geometrical constraint \eqref{gc} as

\begin{equation}\label{gcI}
	w=\frac{2}{\sqrt{\s}}\cos\beta\,{\cal L}^+(\alpha,v)
	\,,
\end{equation}
and the length of the rectangle and the energy of the configuration as 

\begin{equation}\label{EIl}
	\ell^{\,\i}
	=
	\tan\beta\, w+
	\frac{1}{\sqrt{\s}}{\cal L}(\bar\lambda,v)
	\,,\qquad
	E^{\,\i}=\g\sqrt{\s}\Bigl(4{\cal E}^+(\alpha,v)+{\cal E}(\bar\lambda,v)+6\k\frac{\ep^{-2v}}{\sqrt{v}}
\Bigr)+4c
\,.
\end{equation}
In the last equation we used \eqref{baryon-v}. Given the tangent angles as functions of $v$, the function $E(\ell)$ can be written in parametric form as $\ell=\ell^{\,\i}(v)$ and $E=E^{\,\i}(v)$.

Let us first discuss the large-$\ell$ limit, which corresponds to $\bar\lambda\to 1$. In this limit, string $3$ becomes infinitely long, while the other strings remain finite. The tangent angle $\bar\alpha$ can be expressed in terms of $v$ as $\cos\bar\alpha=v\ep^{1-v}$ \cite{a4Q}. On the other hand, all the angles can be expressed in terms of $v$ and $w$ using Eqs.\eqref{fbec} and \eqref{gcI}. Combining the two expressions for $\bar\alpha$ leads to the equation $\bar\alpha(v,w)=\arccos v\ep^{1-v}$, whose solution $\vo$ provides an upper bound on $v$. Clearly, $\vo$ depends on $w$. 

Before trying to understand the details, two comments are in order. First, this is not the true IR limit, but rather the diquark limit. One can show that, in this limit, the points $Y$ and $\bar Y$ do not coincide with the Steiner points of the rectangle as $\beta\not=\frac{\pi}{6}$. Second, there is an upper bound on $w$ for which this limit exists, as we explain below.  

Using Eqs.\eqref{iLy=1} and \eqref{Ey=1}, we obtain the singular terms in $\ell^{\,\i}$ and $E^{\,\i}$

\begin{equation}\label{EIdi}
\ell^{\,\i}
=-\frac{1}{\sqrt{\s}}\ln(1-\bar\lambda)\,+O(1)\,,
\qquad
E^{\,\i}=-\g\ep\sqrt{\s}\ln(1-\bar\lambda)\,+O(1)
\,.
\end{equation}
From this it follows 

\begin{equation}\label{EIdi2}
E^{\,\i}=\sigma\ell\,+O(1)
\,,
\end{equation}
as expected in the diquark limit (see \eqref{EQQb-large}).

To compute the next term, consider

\begin{equation}\label{EIdi3}
E^{\,\i}-\sigma\ell
=
\g\sqrt{\s}\Bigl(4{\cal E}^+(\alpha,v)-2\ep\sin\beta{\cal L}^+(\alpha,v)+{\cal E}(\bar\lambda,v)-\ep{\cal L}(\bar\lambda,v)+6\k\frac{\ep^{-2v}}{\sqrt{v}}\Bigr)
+4c
\,.
\end{equation}
Taking the limit $\bar\lambda\rightarrow 1$ ($v\to\vo$) and using formula \eqref{JLE}, we arrive at 

\begin{equation}\label{EIdi4}
E^{\,\i}-\sigma\ell
=
\g\sqrt{\s}\Bigl(4{\cal E}^+(\alpha,\vo)-2\ep\sin\beta{\cal L}^+(\alpha,\vo)-{\cal J}(\vo)
+6\k\frac{\ep^{-2\vo}}{\sqrt{\vo}}\Bigr)
+4c+o(1)
\,.
\end{equation}
Here $\alpha=\alpha(\vo)=\arcsin\oh\bigl(\sqrt{1-\vo^2\ep^{2(1-\vo)}}+3\k(1+4\vo)\ep^{-3\vo}\bigr)$ and $\beta=\beta(\vo)=\arcsin \vo \ep^{1-\vo}/2\cos\alpha$. Finally, we obtain

\begin{equation}\label{EIdi5}
E^{\,\i}=
\sigma\ell
+
C^{\,\i}+o(1)
\,,
\qquad\text{with}\qquad
C^{\,\i}=\g\sqrt{\s}\Bigl(4{\cal E}^+(\alpha,\vo)-2\ep\sin\beta{\cal L}^+(\alpha,\vo)-{\cal J}(\vo)
+6\k\frac{\ep^{-2\vo}}{\sqrt{\vo}}\Bigr)
+4c
\,.
\end{equation}
It is worth noting that the constant term $C^{\,\i}$ is not universal, as it depends on $w$. 

It is instructive to consider the double limit in which $w\to 0$ as well. We therefore relax the geometrical constraint and let $v\to 0$ together with $\bar \lambda\to 1$. To analyze the behavior of the parametric equations in a small neighborhood of the point $(0,1)$ in the two-dimensional parameter space $(v,\bar\lambda)$, we choose the path $v=\varepsilon$ and $1-\bar\lambda=\varepsilon$. This choice allows us to obtain expansions in powers of $\varepsilon$ and to control path-dependent terms. A straightforward, though somewhat tedious, calculation gives

\begin{equation}\label{diq-limit}
\begin{split}
w=&\sqrt{\frac{v}{\s}}\Bigl(w_0+w_1 v\Bigr)+o(\epsilon^{\frac{3}{2}})
\,,\quad
\ell^{\,\i}=\frac{1}{\sqrt{\s}}\Bigl(-\ln(1-\bar\lambda)+\ell_0\Bigr)+o(\epsilon^{\oh})
\,,\\
E^{\,\i}=&\g\sqrt{\s}\Bigl(-\ep\ln(1-\bar\lambda)+{\cal E}_0\Bigr)
+
4c
+
2\g\sqrt{\frac{\s}{v}}\Bigl(E_0+E_1 v\Bigr)+o(\epsilon^{\oh})
\,,
\end{split}
\end{equation}
where 
\begin{gather}
w_0=\frac{1}{2}\xi^{-\frac{1}{2}}B\bigl(\xi^2;\tfrac{3}{4},\tfrac{1}{2}\bigr)\,,
\qquad
w_1=\frac{1}{2}\xi^{-\frac{3}{2}}\Bigl(
 \Bigl(2\xi+\frac{3}{4}\frac{\k-1}{\xi}\Bigr)B\bigl(\xi^2;\tfrac{3}{4},-\tfrac{1}{2}\bigr)
 -
B\bigl(\xi^2;\tfrac{5}{4},-\tfrac{1}{2}\bigr)\Bigr)\,,
\label{lLE-dq2}\\
E_0=1+3\k+\frac{1}{2}\xi^{\frac{1}{2}}B\bigl(\xi^2;-\tfrac{1}{4},\tfrac{1}{2}\bigr)\,,
\qquad
E_1=-1-6\k+\frac{1}{2}\xi^{-\frac{1}{2}}B\bigl(\xi^2;\tfrac{1}{4},\tfrac{1}{2}\bigr)
+\xi\,w_1 \,,
\label{lLE-dq3}
\end{gather}
and 
\begin{equation}\label{lLE-dq4}
\ell_0=
\int^1_0 du\biggl(2\Bigl(u^{-4}\ep^{2(u^2-1)}-1\Bigr)^{-\frac{1}{2}}
-\frac{1}{1-u}\biggr)\,,
\quad
{\cal E}_0=2\int^1_0 \frac{du}{u^2}\biggl(\ep^{u^2}\Bigl(1-u^4\ep^{2(1-u^2)}\Bigr)^{-\frac{1}{2}}
-1-u^2-\frac{\ep\, u^2}{2(1-u)}\biggr)
\,.
\end{equation}
Here $\xi=\cos\alpha\vert_{v=0}=\frac{\sqrt{3}}{2}\sqrt{1-2\k-3\k^2}\approx 0.938$. When no ambiguity arises, we indicate the dependence on $v$ and $\bar\lambda$ rather than on $\varepsilon$. Note that the integrals in \eqref{lLE-dq4} can be evaluated numerically, with the result $\ell_0\approx -0.710$ and ${\cal E}_0\approx-3.431$. One can then express $E^{\,\i}$ in terms of $\ell$ and $w$. The result is given by Eq.\eqref{diquarkEc} and is consistent with quark-diquark symmetry \cite{wise}.

Formally, the small-$\ell$ limit does not exist unless $w=0$. The geometrical constraint prevents one from taking the limit $v\to 0$. Instead, configuration (I) transforms into another configuration with the negative tangent angle $\alpha$, which we call configuration (II). This occurs at $v=\vz$, where $\vz$ is a solution of the equation $\alpha(v,w)=0$. One may verify, at least numerically, that such a solution exists for any $w$ not exceeding a limiting value $\w$. Moreover, it depends on $w$ and satisfies the inequality $\vz\leq\vo$. Hence, $\vz$ provides a lower bound on $v$.

The width $w$ reaches its limiting value when the interval $[\vz, \vo]$ shrinks into a point. This occurs at $w=\w$, where $\w$ is a solution of the equation $\vz(w)=\vo(w)$. In practice, it is more convenient to first determine $v_{\w}$, which corresponds to $\w$. After substituting $\alpha=0$ and $\bar\alpha=\arccos v\ep^{1-v}$ into the second equation of \eqref{fbec}, we obtain the equation 

\begin{equation}\label{v1c}
	1-9\k^2(1+4v)^2\ep^{-6v}-v^2\ep^{2(1-v)}=0
	\,,
\end{equation} 
whose solution is $v_{\w}$. Given $v_\w$, the geometrical constraint yields

\begin{equation}\label{w}
	\w=\frac{1}{\sqrt{\s}}\sqrt{4-v^2_{\w}\ep^{2(1-v_{\w})}}\,{\cal L}^+(v_{\w})
	\,.
\end{equation}
Note that, for the parameter set of Sec.II, a simple calculation gives $\w=0.779\,\text{fm}$.

In summary, configuration (I) exists for $w\leq\w$. In this case its energy is given parametrically by $\ell=\ell^{\,\i}(v)$ and $E^{\,\i}=E^{\,\i}(v)$, with the parameter running from $\vz$ to $\vo$.

\subsection{Configuration (II)}

Since this configuration is governed by the same action as configuration (I), the force balance equations \eqref{fbec} hold. As explained in Appendix B of \cite{a4Q}, the expressions for the geometrical constraint, the length, and the energy can be obtained by replacing ${\cal L}^+$ and ${\cal E}^+$ in Eqs.\eqref{gcI} and \eqref{EIl} with ${\cal L}^-$ and ${\cal E}^-$, corresponding to negative $\alpha$. This gives 

\begin{equation}\label{gcII}
	w=\frac{2}{\sqrt{\s}}\cos\beta\,{\cal L}^-(\lambda,v)
	\,
\end{equation}
and 
\begin{equation}\label{ElII}
	\ell^{\,\ii}=
	\tan\beta\, w
	+\frac{1}{\sqrt{\s}}{\cal L}(\bar\lambda,v)
	\,,\qquad
	E^{\,\ii}=\g\sqrt{\s}\Bigl(4{\cal E}^-(\lambda,v)+{\cal E}(\bar\lambda,v)+6\k\frac{\ep^{-2v}}{\sqrt{v}}
\Bigr)+4c
\,,
\end{equation}
with $\lambda=-\text{ProductLog}(-v\ep^{-v}/\cos\alpha)$. As before, the angles can be expressed in terms of $v$, and the energy as a function of the rectangle's length can be written parametrically as $\ell=\ell^{\,\ii}(v)$ and $E=E^{\,\ii}(v)$. 

The only issue with this expression is that a domain of the parameter $v$ is not specified. To determine it, we first consider the case $w<\w$. As we saw in the previous section, configuration (I) transforms into configuration (II) at $v=\vz$. The question then arises as to what happens with configuration (II) for smaller values of $v$. The answer to that is that it continues to exist down to $\vcd$, which is a solution to the equation $\bar\alpha(v,w)=0$ and corresponds to a transition to the pinched tetraquark configuration (d). One can verify numerically that such a solution exists in the interval $[0,1]$. In addition, it depends on $w$ and satisfies the inequality $\vcd<\vz$. The latter implies that for $w\leq\w$ the parameter $v$ varies from $\vcd$ to $\vz$. Thus, in this case configuration (II) is relevant only for intermediate lengths, and it makes no sense to consider the small- and large-$\ell$ limits. Putting all of this  together, for $w<\w$ the tetraquark configuration is described in terms of both basic configurations. Its energy is given parametrically by the piecewise functions arising from configurations (I) and (II), see \eqref{Elc1}.

The description simplifies for $w\geq\w$, where configuration (I) becomes irrelevant because the interval $[\vz,\vo]$ shrinks into a point. This allows us to consider the large-$\ell$ limit, which can be analyzed similarly to that of configuration (I). Again, only string $3$ becomes infinitely long at finite $w$, so that the corresponding tangent angle is given by $\cos\bar\alpha=v\ep^{1-v}$. After expressing this angle in terms of $v$ from Eqs.\eqref{fbec} and \eqref{gcII}, we obtain the equation $\bar\alpha(v,w)=\arccos v\ep^{1-v}$, whose solution is $\vo$. As before, $\vo$ depends on $w$. 

The leading contributions to $\ell^{\,\ii}$ and $E^{\,\ii}$ as $\bar\lambda\to 1\,(v\to\vo)$ are given by the same expressions as those on the right hand side of \eqref{EIdi}. Therefore, $E^{\,\ii}$ also grows linearly for large $\ell$. To find the constant term, we consider the difference between $E^{\,\ii}$ and $\sigma\ell$ 

\begin{equation}\label{EIIdi3}
E^{\,\ii}-\sigma\ell
=
\g\sqrt{\s}\Bigl(4{\cal E}^-(\lambda,v)-2\ep\sin\beta{\cal L}^-(\lambda,v)+{\cal E}(\bar\lambda,v)-\ep{\cal L}(\bar\lambda,v)+6\k\frac{\ep^{-2v}}{\sqrt{v}}\Bigr)
+4c
\,
\end{equation}
and then take the limit $\bar\lambda\to 1$. We thus obtain

\begin{equation}\label{EIIdi4}
E^{\,\ii}-\sigma\ell
=
\g\sqrt{\s}\Bigl(4{\cal E}^-(\lambda,\vo)-2\ep\sin\beta{\cal L}^-(\lambda,\vo)
-{\cal J}(\vo)+6\k\frac{\ep^{-2\vo}}{\sqrt{\vo}}\Bigr)
+4c
\,, 
\end{equation}
where $\beta=\beta(\vo)$ and $\lambda=-\text{ProductLog}(-\vo\ep^{-\vo}/\cos\alpha(\vo))$. Finally, 

\begin{equation}\label{EIIdi5}
E^{\,\ii}=
\sigma\ell
+
C^{\,\ii}+o(1)
\,,
\quad
C^{\,\ii}=\g\sqrt{\s}\Bigl(4{\cal E}^-(\lambda,\vo)-2\ep\sin\beta{\cal L}^-(\lambda,\vo)-{\cal J}(\vo)
+6\k\frac{\ep^{-2\vo}}{\sqrt{\vo}}\Bigr)
+4c
\,.
\end{equation}
The constant term is again $w$-dependent. 

At this point a couple of brief remarks are in order. First, although $\ell$ tends to infinity, the points $Y$ and $\bar Y$ do not coincide with the Steiner points as long as $w$ is finite. Second, in the true IR limit where $w$ also tends to infinity, the external strings become infinitely long, with the tangent angles given by $\cos\alpha=v\ep^{1-v}$. In this case, $\vo$ is determined by the equation 

\begin{equation}\label{w-infty}
	1-\k^2(1+4v)^2\ep^{-6v}-v^2\ep^{2(1-v)}=0
	\,.
\end{equation}
It then follows from the first equation in \eqref{fbec} that $\beta=\frac{\pi}{6}$, and therefore the points $Y$ and $\bar Y$ are the Steiner points of the rectangle. 

Thus, for $w\geq\w$, the energy of configuration (II) can be written parametrically as $\ell=\ell^{\,\ii}(v)$ and $E=E^{\,\ii}(v)$, with the parameter varying from $\vcd$ to $\vo$. In this case, the tetraquark configuration is described only by the basic configuration (II), see \eqref{Elc2}.


\end{document}